\documentclass[12pt,hyper,notoc]{JHEP3}

\title{Primordial Magnetic Field Amplification from Turbulent Reheating}
\author{Esteban Calzetta\\Depto. de F\'{\i}sica, FCEyN-UBA and IFIBA-CONICET, 
Cdad. Universitaria, Buenos Aires, Argentina\\ Email: 
\email{calzetta@df.uba.ar}}
\author{Alejandra Kandus\\ LATO - DCET - UESC. Rodovia Ilh\'{e}us-Itabuna, km 16 
s/n, CEP: 45662-900,
Salobrinho, Ilh\'{e}us-BA, Brazil \\ E-mail: \email{kandus@uesc.br}}

\usepackage{amsfonts}
\usepackage{amsmath}
\usepackage{amssymb}
\usepackage{epsfig}

\abstract{
We analyze the possibility of primordial magnetic field amplification by a
stochastic large scale kinematic dynamo during reheating. We consider a
charged scalar field minimally coupled to gravity. During inflation this
field is assumed to be in its vacuum state. At the transition to reheating
the state of the field changes to a many particle/anti-particle state. We
characterize that state as a fluid flow of zero mean velocity but with a
stochastic velocity field. We compute the scale-dependent Reynolds number $%
Re(k)$, and the characteristic times for decay of turbulence, $t_{d}$ and
pair annihilation $t_{a}$, finding $t_{a}\ll t_{d}$. We calculate the rms
value of the kinetic helicity of the flow over a scale $\mathcal{L}$ and
show that it does not vanish. We use this result to estimate the
amplification factor of a seed field from the stochastic kinematic dynamo
equations. Although this effect is weak, it shows that the evolution of the
cosmic magnetic field from reheating to galaxy formation may well be more
complex than as dictated by simple flux freezing.}


\begin{document}

\section{ Introduction}

The question of the origin of large scale magnetic fields that permeate
almost all structures of the universe is one of the most challenging areas
of research in astrophysics. None of the main lines of investigation, namely
primordial origin or in situ generation, succeeded up to now to explain both
the intensity and the topology of the large scale fields. Local generation
mechanisms are mainly based on seed field generation by, e.g., a local
battery, amplified by a turbulent dynamo in the interstellar or
intergalactic medium (see \cite{bran-sub-05} and references therein). The
primordial origin hypothesis, on the other hand, considers that at least the
seed field is generated at some early epoch (inflation, reheating or
radiation dominance), and is amplified by flux conservation and/or turbulent
dynamo action during gravitational collapse from $z\approx 100$ on \cite%
{CalKan99}. The seed field must be quite intense for gravitational collapse
to produce the detected intensities, and the turbulent dynamo must operate
almost since the birth of the galaxy, i.e., during most of the matter
dominated era. The recent detection of regular fields in high redshift
quasars \cite{wolfe-08}, \cite{kronberg-nat-08}, \cite{kronberg-apj-08}
however may challenge the in situ generation, or at least the dynamo
mechanism in the form we understand it today, favoring the primordial origin
of the fields.

Two obstacles must be overcome by a successful primordial generation
mechanism: breaking the conformal symmetry of a massless gauge field in a
spatially flat universe and building a large coherence length. Sub-horizon
processes, like phase transitions \cite{hogan-83}, \cite{qls-89}, \cite%
{co-94}, \cite{soj-97}, in general produce intense fields, but of very small
coherence length (see Refs \cite{gra-rub-01} and \cite{widrow-02} for
reviews of different magnetogenesis mechanisms). The inflationary epoch of
the universe (if ever existed) offers a suitable scenario for large scale
field generation as in it sub-horizon scales naturally become super-horizon.
Several mechanisms were considered along the years in which conformal
invariance is broken either by coupling the magnetic field to curvature in
different ways or addressing non-linear electrodynamics \cite{turn-wid-88}, 
\cite{mazz-sped-95}, \cite{tsa-kan-05}, \cite{kunze-08}. In general the
fields produced are extremely weak, or of marginal intensity, to seed
subsequent amplification processes. The reheating period has also been
studied as a magnetogenesis scenario (\cite{cal-kan-mazz}, \cite%
{kan-cal-mazz-wag}, \cite{giov-shap}, \cite{cal-kan-02}, \cite{maroto-01})
but in all scenarios considered so far the obtained fields are too weak to
be of astrophysical interest.

Confronted with this situation one wonders if it is possible to have a
pre-amplification (or perhaps full amplification) of a seed field created by
one of the above mentioned mechanisms already in the early universe. In this
sense the reheating epoch offers a good prospect, as it is a period where
highly non-linear and out of equilibrium processes take place \cite%
{CalHu08,khleb-tkach-96,khleb-tkach-97,ko-li-sta-96,ko-li-sta-97,fin-bran-99,fin-bran-00,feld-tkach-08,feld-gbell-etal-01,feld-kof-01,gra-cal-02,
je-le-ma-10}. This possibility was explored for the first time some years 
ago by Finelli and Gruppuso \cite{FiGru-01} and by Bassett et al.
\cite{BaPoTsuVi-01}. In Ref. \cite{FiGru-01} it is analyzed the amplification 
of a pre-existing  magnetic field by parametric resonance during the 
oscillatory regime of a scalar field to which the magnetic field is 
coupled. In Ref. \cite{BaPoTsuVi-01} the amplification during preheating 
is studied considering several different models.
Another possibility for such pre-amplification process, and that
will be investigated in this paper, could be
the operation of a turbulent large scale dynamo \cite{moffatt}, \cite%
{dan-pab-1}, \cite{bran-sub-05}, \cite{mininni-07}, similar to the one that
acts in the interstellar plasma.

That the matter fields in reheating can be turbulent was pointed out in
Refs. \cite{khleb-tkach-96}, \cite{feld-tkach-08}, \cite{feld-gbell-etal-01}%
, \cite{feld-kof-01} (see \cite{gra-cal-02} for a theoretical analysis of
turbulent reheating). A dynamo requires the presence of a plasma. As the
inflaton is a gauge singlet, it will not decay directly into charged
species. Therefore to have a plasma we must consider an extra, charged,
field. The mechanism by which the plasma is created is particle creation
during the transition from inflation to reheating \cite%
{birrel-davies,MukWin07,CalHu08,ParTom09}.

Suppose that the charged species in question was in its vacuum state during
inflation. The created particles will generate stochastic currents that on
one hand induce a seed field \cite{cal-kan-mazz}, \cite{giov-shap} and on
the other may constitute the turbulent plasma we are looking for. Creation
of spin 1/2 particles such as electrons is suppressed by conformal
invariance at the high energies prevailing during inflation \cite%
{cal-kan-mazz}, so the charged species must be a scalar. Suitable candidates
can be found in supersymmetric extensions of the standard model \cite%
{kan-cal-mazz-wag}.

The simplest model for a turbulent large scale dynamo is driven by flow
velocities and does not take into account the back-reaction of the amplified
fields. It is known as a \textsl{kinematic dynamo} \cite{moffatt}, \cite%
{dan-pab-1}, \cite{bran-sub-05}, \cite{mininni-07}. The sufficient condition
for it to be operational is the flow to be helical, i.e., that the volume
average of the scalar product of the vorticity (curl of the velocity) and
the velocity, known as \textsl{kinetic helicity} \cite{lesieur}, does not
vanish \cite{bran-big-sub-01}, \cite{min-go-mah-05}. Of course this
approximation (the neglect of the back reaction of the induced field) is
valid for weak magnetic fields and/or very short times of operation.
Mathematically speaking, the equation for that dynamo can be written as $%
\partial \overline{B^{i}}/\partial t\simeq -t_{corr}\mathcal{H}_{c}\epsilon
_{ijk}\partial B^{k}/\partial x^{j}$, where $\overline{B^{i}}$ is the large
scale field (or mean field), $\mathcal{H}_{c}$ the kinetic helicity and
$t_{corr}$ a correlation time. If $\epsilon _{ijk}\partial B^{k}/\partial x^{j}
\sim \overline{B^{i}}/L$, with $L
$ the coherence length of the field, then we can estimate $\overline{B^{i}}%
\left( t\right) \sim \overline{B_{0}^{i}}\left( 0\right) \exp \left(
-t_{corr}\mathcal{H}_{c}t/L\right) $.

In this paper we shall investigate the possibility of a dynamo action during
reheating. We assume the existence of a charged scalar field, minimally
coupled to gravity, that is in its vacuum state during inflation. To
simplify the analysis we consider de Sitter inflation and thus a de Sitter
invariant vacuum for the field \cite{allen}. As mentioned above, when the
transition from inflation to reheating takes place, the scalar field is
amplified, and stochastic currents are generated. The characterization of
these particles as a fluid is straightforward. The hydrodynamic energy and
pressure are determined by matching the expectation value of the
energy-momentum tensor of the scalar field to that of a perfect fluid at
rest.

The fluid has a stochastic Gaussian velocity, which is found by matching the
self-correlation of the $0i$ components of the energy momentum tensor of the
fluid to the symmetric expectation value of the corresponding operator for
the field. Finally, the viscosity of the fluid is found by assuming that it
is close to saturate the Kovtun, Son and Starinets bound \cite{son-1}. While
initially derived from consideration of the AdS/CFT correspondence, the fact
that a similar bound seems to hold for the strongly coupled quark gluon
plasma \cite{LuzRom09} suggests that this bound is a good description of
field theories in general.

We characterize the turbulence by finding the momentum dependent Reynolds
number $Re\left( k\right) $. As for the magnetic Reynolds number, $R_{m}$ we
do not need to estimate it because we are interested in the kinematic
regime, where magnetic fields are too weak to backreact on the flow. As
there are no stirring forces, turbulence will decay eventually. We calculate
the decay time of the turbulence for each mode, $t_{d}\left( k\right) $. On
the other hand, the fluid is made of particles and antiparticles, which are
liable to annihilate. We also estimate the characteristic time for pair
annihilation, $t_{a}\left( k\right) $, finding that $t_{a}\left( k\right)
<t_{d}\left( k\right) $, i.e., the fluid annihilates before turbulence
decays. This fact allows us to consider that the turbulence is stationary in
the interval $0\leq t\leq t_{a} $.

The non-trivial result of our paper is that the rms value of the kinetic
helicity of the fluid, $\mathcal{H}_{c}$, is not zero. This proves that the
kinematic dynamo action mentioned above is indeed possible. The key
ingredient to have $\mathcal{H}_{c}\neq 0$, is the fact that the plasma is
made up of two scalar fields, $\Phi $ and $\Phi ^{\dagger }$. We then
estimate the amplification factor of the induced field based on the dynamo
equation written above, i.e., $\sim \exp \left[ \mathcal{H}_{c}t_{corr}/L\right]
$.

We work with signature $\left( -,+,+,+\right) $ and with natural units,
i.e., $\hbar =c=k_{B}=1$. Greek indexes denote space-time coordinates while
latin indexes refer to spatial coordinates. To simplify our analysis we
shall consider de Sitter inflation, and define dimensionless variables and
fields as $x^{i}=Hr^{i}$, $\tau =Ht$, and $\Phi =H^{-1}\Psi $, where $H$ is
the Hubble constant that characterizes the de Sitter phase. The mass of
the scalar field will combine with $H$ to produce the dimensionless mass
parameter, $m/H$.
In section I we
make a brief review of dynamo theory. Section II is devoted to the fluid
description of a quantum field. In it we find the velocity correlation
function and velocity spectrum as well as the kinetic helicity correlation
function. In section III\ we characterize the turbulence by finding the
Reynolds number, $Re\left( k\right) $, and the characteristic times $t_{d}$
and $t_{a}$. In section IV we find the amplification factor for the magnetic
field. Finally in section V we summarize our conclusions. The bulk of the
calculations that lead to these results are shown in the four appendices.

\section{Basics of mean field dynamo theory}

In this section we briefly sketch the so called \textsl{first order
smoothing approximation} (FOSA) approach to the theory of mean field dynamo.
We refer the reader to Refs \cite{moffatt}, \cite{bran-sub-05}, \cite%
{rad-rhein-07} and references therein for details. In FOSA, purely
hydrodynamic turbulence is considered, ignoring higher than second order
correlations in the fluctuating velocity field $u^{i}$. This approach is
suitable for short times and for magnetic fields that are weak enough to
neglect their backreaction on the turbulent flow. In short, fields are
divided in mean and fluctuating components, as%
\begin{equation}
B^{i}=\overline{B^{i}}+b^{i};\quad U^{i}=\overline{U^{i}}+\mathtt{v}^{i}
\label{mft-1}
\end{equation}%
where overbar denotes volume average, and that is assumed that satisfies 
Reynolds rules \cite{mccomb}.
In the case that $\overline{U^{i}}=0$, the mean magnetic field satisfies the
equation%
\begin{equation}
\frac{\partial \overline{B^{i}}}{\partial t}=\epsilon ^{ijk}\frac{\partial 
\overline{\mathcal{E}^{k}}}{\partial x^{j}}+\frac{\eta _{0}}{a^{2}\left(
\tau \right) }\nabla ^{2}\overline{B^{i}}  \label{mft-2}
\end{equation}%
The important quantity here is the mean electromotive force, $\overline{%
\mathcal{E}}$, given by%
\begin{equation}
\overline{\mathcal{E}^{i}}=\epsilon _{ijk}\overline{\mathtt{v}^{j}b^{k}}
\label{mft-3}
\end{equation}%
If $\overline{B^{i}}$ is sufficiently weak and regular, $\mathcal{E}^{i}$
can be expanded as \cite{bran-sub-05} 
\begin{equation}
\overline{\mathcal{E}^{i}}=\int_{0}^{t}\left[ \alpha _{ip}\left( t,t^{\prime
}\right) \overline{B^{p}}\left( t^{\prime }\right) +\beta _{ikp}\left(
t,t^{\prime }\right) \frac{\partial \overline{B^{p}}\left( t^{\prime
}\right) }{\partial x_{k}}\right] dt^{\prime }  \label{mft-4}
\end{equation}%
with $\alpha _{ip}=\epsilon _{ijk}\overline{\mathtt{v}_{j}\left( t\right)
\partial \mathtt{v}_{k}\left( t^{\prime }\right) /\partial x_{p}}$ and $%
\beta _{ikp}\left( t,t^{\prime }\right) =\epsilon _{ikp}\overline{\mathtt{v}%
_{l}\left( t\right) \mathtt{v}_{p}\left( t^{\prime }\right) }$. Under the
hypothesis of local homogeneity and isotropy, the tensors in the integrand
must be proportional to $\delta _{ip}$ and $\epsilon _{ikp}$ respectively
eq. (\ref{mft-4}) can be written as 
\begin{equation}
\overline{\mathcal{E}^{i}}=\int_{0}^{t}\left[ \alpha \left( t-t^{\prime
}\right) \overline{B^{i}}\left( t^{\prime }\right) +\beta \left( t-t^{\prime
}\right) \overline{J^{i}}\left( t^{\prime }\right) \right] dt^{\prime }
\label{mft-5}
\end{equation}%
where $\alpha \left( t-t^{\prime }\right) =-\left( 1/3\right) \overline{%
\mathtt{v}\left( t\right) \cdot \mathtt{w}\left( t^{\prime }\right) }$, $%
\mathtt{w}^{i}=\epsilon ^{ijk}\partial _{j}\mathtt{u}_{k}$ being the
vorticity of the velocity fluctuations; $\beta \left( t-t^{\prime }\right)
=\left( 1/3\right) \overline{\mathtt{v}\left( t\right) \cdot \mathtt{v}%
\left( t^{\prime }\right) }$ and $\overline{J^{i}}\left( t\right) $ the mean
electric current. If besides it is assumed that $\overline{B^{i}}\left(
t\right) $ is a slowly varying function of time, then eq. (\ref{mft-5})
turns into%
\begin{equation}
\overline{\mathcal{E}^{i}}=\alpha \overline{B^{i}}-\beta \overline{J^{i}}
\label{mft-5b}
\end{equation}%
with%
\begin{equation}
\alpha =-\frac{1}{3}\int_{0}^{t}\overline{\mathtt{v}\left( t\right) \cdot 
\mathtt{w}\left( t^{\prime }\right) }dt^{\prime }\approx -\frac{1}{3}t_{corr}%
\overline{\mathtt{v}\left( t\right) \cdot \mathtt{w}\left( t\right) }
\label{mft-6}
\end{equation}%
and%
\begin{equation}
\beta =\frac{1}{3}\int_{0}^{t}\overline{\mathtt{v}\left( t\right) \cdot 
\mathtt{v}\left( t^{\prime }\right) }dt^{\prime }\approx \frac{1}{3}t_{corr}%
\overline{\mathtt{v}^{2}\left( t\right) }  \label{mft-7}
\end{equation}%
with $t_{corr}$ the correlation time. The last approximation in eqs. (\ref%
{mft-6}) and (\ref{mft-7}) is known as the \textquotedblleft $\tau $
approximation\textquotedblright\ \cite{bran-sub-05}. Observe that $\alpha $
is minus the \textsl{kinetic helicity}, $\mathcal{H}_{c}$ of the flow \cite%
{lesieur}. A non-null value of this quantity indicates that the flow lacks
mirror symmetry. This is a sufficient condition for dynamo action \cite%
{moffatt,dan-pab-1,mininni-07}. If $\overline{B^{p}}$ is smoothly varying,
then the dominant term in eq. (\ref{mft-5b}) is the first one, and eq. (\ref%
{mft-2}) can be written as%
\begin{equation}
\frac{\partial \overline{B^{i}}}{\partial t}\simeq -\frac{1}{3}t_{corr}%
\mathcal{H}_{c}\epsilon _{ijk}\frac{\partial B^{k}}{\partial x^{j}}
\label{mft-8}
\end{equation}%
Taking $\epsilon _{ijk}\partial \overline{B^{k}}/\partial x^{j}\sim 
\overline{B^{i}}/\mathtt{L}$, with $\mathtt{L}$ the scale of coherence of
the mean field, eq. (\ref{mft-8}) can be directly integrated for short
times. We shall show below that in our case $\mathcal{H}_c$ is a Gaussian variable of
zero mean value and known variance $\Sigma_{\mathcal{H}_c}$. Taking the ensemble
average over all possible realizations of $\mathcal{H}_c$ we obtain that the
mean magnetic field is
\begin{equation}
\overline{B^{i}}\left( t\right)
\sim \overline{B_{0}^{i}}\exp \left( \frac{1}{2}\left\langle 
\frac{1}{9}t_{corr}^{2}\frac{\Sigma_{\mathcal{H}_c}^2}{\mathtt{L}^{2}}%
t^2\right\rangle \right)   \label{mft-9}
\end{equation}%
with $\overline{B_{0}^{i}}$ the initial value of the field. Our task in the
next sections is to characterize the system of cosmological created scalar
particles as a turbulent flow, and investigate if it has a non-zero kinetic
helicity.

\section{Fluid description of charged quantum scalar fields}

Consider a charged scalar field, $\left( \Phi ,\Phi ^{\dagger }\right) $,
minimally coupled to gravity in a spatially flat Friedmann-Robertson-Walker
universe, described by the line element $dS^{2}=-dt^{2}+a^{2}\left( t\right)
\left( dx^{2}+dy^{2}+dz^{2}\right) $, with $a\left( t\right) $ the expansion
factor. We assume that the e.m. field is so weak that it can be neglected
throughout. The action of the field reads%
\begin{equation}
S=-\frac{1}{2}\int d^{4}x\sqrt{-g}\left[ g^{\alpha \beta }\partial _{\alpha
}\Phi \partial _{\beta }\Phi ^{\dagger }+\frac{m^{2}}{H^{2}}\Phi \Phi
^{\dagger }\right]  \label{a}
\end{equation}%
with $g^{\mu \nu }$ the spacetime metric, $m/H$ 
the dimensionless mass parameter of the field and $H=%
\dot{a}\left( \tau \right) /a\left( \tau \right) $ the Hubble constant
during inflation. Throughout the paper we consider $m/H\ll 1$ (see e.g. \cite%
{kan-cal-mazz-wag}). The stress energy tensor is given by 
\begin{equation}
T_{\Phi }^{\mu \nu }=\frac{-2}{\sqrt{-g}}\frac{\delta S}{\delta g^{\mu \nu }}
\label{b}
\end{equation}%
Explicitly 
\begin{equation}
T_{\Phi }^{\mu \nu }=H^{4}\left[ \partial ^{\mu }\Phi \partial ^{\nu }\Phi
^{\dagger }-\frac{1}{2}g^{\mu \nu }\partial _{\alpha }\Phi \partial ^{\alpha
}\Phi ^{\dagger }-\frac{1}{2}g^{\mu \nu }\frac{m^{2}}{H^{2}}\Phi \Phi
^{\dagger }\right]  \label{c}
\end{equation}%
The electric current density is%
\begin{equation}
J_{\Phi }^{\mu }=ieH^{3}\left[ \Phi \ \partial ^{\mu }\Phi ^{\dagger
}-\partial ^{\mu }\Phi \ \Phi ^{\dagger }\right]  \label{e}
\end{equation}%
We identify $T_{\Phi }^{\mu \nu }$ with the stress energy tensor of a two
fluid system. One fluid corresponds to the positively charged scalar
particles, and the other to the negatively charged anti-particles.
Analogously we identify $J_{\Phi }^{\mu }$ with the electric current of the
two fluid system. To this purpose we define the four velocity of the flow as
usual, i.e.,%
\begin{equation}
u^{\mu }\equiv \gamma \left( U^{\mu }+\mathtt{v}^{\mu }\right)
\label{4-vel-a}
\end{equation}%
with $U^{\mu }=\left( 1,0,0,0\right) $ the four velocity of the fiducial
observers at rest with respect to the radiation field produced by the decay
of the inflaton. We define the projector onto the surface orthogonal to the
world lines of fiducial observers in the usual way, i.e., $h^{\mu \nu
}=g^{\mu \nu }+U^{\mu }U^{\nu }$, so $h^{\mu \nu }U_{\nu }=0$ and $h^{\mu
\nu }\mathtt{v}_{\nu }=\mathtt{v}^{\mu }$. We then write%
\begin{equation}
T_{\Phi }^{00}\equiv \left\langle \rho +p\right\rangle \gamma
^{2}U^{0}U^{0}+pg^{00}  \label{f}
\end{equation}%
\begin{equation}
T_{\Phi }^{\left\{ 0i\right\} }\equiv \left\langle \rho +p\right\rangle
\gamma ^{2}U^{0}\mathtt{v}^{i},\quad \mathtt{v}^{i}=\mathtt{v}_{+}^{i}+%
\mathtt{v}_{-}^{i}  \label{g}
\end{equation}%
\begin{equation}
J_{\Phi }^{i}\equiv en\gamma \left( \mathtt{v}_{+}^{i}-\mathtt{v}%
_{-}^{i}\right)  \label{h}
\end{equation}%
where we have symmetrized $T_{\Phi }^{\left\{ 0i\right\} }=\left( T_{\Phi
}^{0i}+T_{\Phi }^{i0}\right) /2$. In eq. (\ref{g}) $\mathtt{v}_{+}^{i}\left( 
\mathtt{v}_{-}^{i}\right) $ is the stochastic velocity of the positively
(negatively) charged species and in eq. (\ref{h}) $n$ is the number density
of particles. In both equations $\gamma $ is the Lorentz factor due to the
(macroscopic) velocity of the fluid measured by the fiducial observers. Our
flow is made up of gravitationally created particles during the transition
between inflation and reheating. As momentum is conserved in the particle
creation process, in the radiation frame both fluids have zero bulk
velocity, thus we can take $\gamma =1$. Therefore $\mathtt{v}_{\pm }^{i}$
are stochastic fluctuations around the zero mean velocity,
that must be characterized through their correlation function.

\subsection{Transition from inflation to reheating: particle creation}

The stochastic velocity $\mathtt{v}^{i}$ introduced at the beginning of this
section is the result of random motions of scalar charges. To understand how
those charges appear we observe that a state detected as vacuum by inflationary
observers will be detected as a many-particle state by comoving observers in
the reheating epoch. Mathematically this is expressed
as follows \cite{CalHu08,birrel-davies,MukWin07,ParTom09}.

From eq. (\ref{a}) we obtain the evolution equation for the scalar field $%
\Phi $, the Klein-Gordon equation, 
\begin{equation}
\left( \frac{\partial ^{2}}{\partial \tau ^{2}}-\nabla ^{2}-\frac{m^{2}}{%
H^{2}}\right) \Phi =0  \label{j}
\end{equation}%
(and an identical equation for $\Phi ^{\dagger }$). The field operators can
be written in terms of creation and annihilation operators as 
\begin{align}
\Phi & =\frac{1}{\left( 2\pi \right) ^{3/2}}\int \frac{d\bar{\kappa}}{%
a^{3}\left( \tau \right) }\left[ \phi _{\kappa }\left( \tau \right) e^{i\bar{%
\kappa}\cdot \bar{x}}a_{\kappa }+\phi _{\kappa }^{\ast }\left( \tau \right)
e^{-i\bar{\kappa}\cdot \bar{x}}b_{\kappa }^{\dagger }\right]   \label{jb} \\
\Phi ^{\dagger }& =\frac{1}{\left( 2\pi \right) ^{3/2}}\int \frac{d\bar{%
\kappa}}{a^{3}\left( \tau \right) }\left[ \phi _{\kappa }\left( \tau \right)
e^{i\bar{\kappa}\cdot \bar{x}}b_{\kappa }+\phi _{\kappa }^{\ast }\left( \tau
\right) e^{-i\bar{\kappa}\cdot \bar{x}}a_{\kappa }^{\dagger }\right] 
\label{jbb}
\end{align}%
where $\kappa = k/H$, is the dimensionless wavenumber, $k$ the physical wavenumber
and $H$ the Hubble constant during inflation.
Replacing in eq. (\ref{j}) we obtain the evolution equation for each mode $%
\phi _{\kappa }$, i.e., 
\begin{equation}
\ddot{\phi}_{\kappa }+\left[ \frac{\kappa ^{2}}{a^{2}\left( \tau \right) }%
+\left( \frac{m}{H}\right) ^{2}-\frac{3}{2}\left( \frac{\ddot{a}}{a}+\frac{%
\dot{a}^{2}}{2a^{2}}\right) \right] \phi _{\kappa }=0.  \label{k-g}
\end{equation}%
Here $\kappa /a\left( \tau \right) $ is the dimensionless physical
wavenumber. For simplicity we consider de Sitter inflation, where an
invariant vacuum for a minimally coupled scalar field exists \cite{allen},
and we assume that the scalar field is initially in this state. Therefore
the positive energy solutions of eq. (\ref{k-g}) are the Hankel functions $%
H_{\nu }^{\left( 1\right) }$ \cite{allen,birrel-davies}, i.e.,%
\begin{equation}
\phi _{\kappa }^{I}\left( \tau \right) =\frac{\sqrt{\pi }}{2}H_{\nu
}^{\left( 1\right) }\left[ \frac{\kappa }{a\left( \tau \right) }\right] 
\label{k-h-a}
\end{equation}%
with $\nu =\sqrt{9/4-m^{2}/H^{2}}$. We follow Refs. \cite%
{kolb-turner,starob-80} to the effect that during the reheating period the
scale factor of the Universe scales as $t^{2/3}$. In this case it is accurate
enough to consider a WKB approximation for the modes, i.e.,%
\begin{equation}
\phi _{\kappa }^{R}\left( \tau \right) =\frac{\exp \left[ -iS_{\kappa
}\left( \tau \right) \right] }{\sqrt{2\Omega _{\kappa }\left( \tau \right) }}
\label{k-h-b}
\end{equation}%
with $dS_{\kappa }\left( \tau \right) /d\tau =\Omega _{\kappa }\left( \tau
\right) =\sqrt{\kappa ^{2}/a^{2}\left( \tau \right) +\left( m/H\right) ^{2}}$%
. After the transition to reheating the commoving observers in the new
geometry see the state of the scalar field as a many-particle state, \cite%
{birrel-davies,MukWin07,CalHu08,ParTom09}. Mathematically this is described
as 
\begin{equation}
\phi _{\kappa }^{I}\left( \tau \right) =\alpha _{\kappa }\phi _{\kappa
}^{R}\left( \tau \right) +\beta _{\kappa }\phi _{\kappa }^{R\ast }\left(
\tau \right)   \label{corr-6}
\end{equation}%
where $\phi _{\kappa }^{I}\left( \tau \right) $ ($\phi _{\kappa }^{R}\left(
\tau \right) $) are the positive frequency solution of the Klein Gordon
equation for inflation (reheating), and $\alpha _{\kappa }$ and $\beta
_{\kappa }$ the so called Bogoliubov coefficients \cite%
{birrel-davies,MukWin07,CalHu08,ParTom09}. If $\beta _{\kappa }\neq 0$, eq. (%
\ref{corr-6}) shows that a positive frequency wave during inflation becomes
a mixture of positive and negative frequency waves during reheating. The
details of the calculation of these coefficients are given in Appendix A,
here we quote the resulting expressions together with the physical
explanation. The number of created particles in modes with $\kappa <1$,
i.e., super-horizon ones, is not sensitive to the details of the transition.
For $\kappa >1$ that number does depend on the transition features. We take into
account this dependence by assuming the most simple form for it, i.e., a
linear transition that lasts a time $\tau _{0}$. The $\beta _{\kappa }$
coefficient for a linear transition reads%
\begin{equation}
\beta _{\kappa }^{\left( s\right) }\simeq -i\left( \frac{9}{16}\right) ^{2}%
\frac{1}{8}\frac{1}{\tau _{0}^{2}}\frac{1}{\kappa ^{6}}e^{i2\tau _{0}S\left[
0\right] }\sin \left( 2\tau _{0}\kappa \right)   \label{bog-sw-0}
\end{equation}%
where $\tau _{0}$ is the duration of the transition from inflation to
reheating. (see Ref. \cite{zab-sas-09} for a similar analysis, though for
cosmological perturbations). As $\kappa >1$, the modes that are most
amplified are those for which $\tau _{0}\kappa $ is small. For these modes $%
\sin \left( 2\tau _{0}\kappa \right) \sim 2\tau _{0}\kappa $. Hence we take%
\begin{equation}
\beta _{\kappa }^{\left( s\right) }\sim -i\left( \frac{9}{16}\right) ^{2}%
\frac{1}{4}\frac{1}{\tau _{0}}\frac{e^{i2\tau _{0}S\left[ 0\right] }}{\kappa
^{5}}  \label{bog-sw}
\end{equation}%
As for $\kappa <1$ details of the transition do not matter, we have the
usual solution from assuming an instantaneous transition at $\tau = 0$. 
\begin{equation}
\beta _{\kappa }^{\left( l\right) }\simeq \frac{i\left( \nu -1\right) \Gamma
\left( \nu -1\right) }{\pi ^{1/2}\Omega _{\kappa }^{1/2}\left( 0\right) }%
\frac{1}{\kappa ^{\nu }}  \label{bog-lw}
\end{equation}%
with $\nu =\sqrt{9/4-m^{2}/H^{2}}$, and $\Omega _{\kappa }\left( 0\right) =%
\sqrt{\kappa ^{2}+m^{2}/H^{2}}$. After the transition an out of equilibrium
plasma is established. It has no bulk velocity with respect to the comoving
observer's rest frame, but due to the random motions of its constituents,
fluctuating velocities do exist.

\subsection{Two Point Velocity Correlation Function}

One way to characterize a system with fluctuating velocities is to give
their spatial two point velocity correlation function \cite{monin-yaglom}.
It is defined as the equal time ensemble average of the product $\mathtt{v}%
^{i}\left( \tau ,\bar{x}\right) \mathtt{v}^{j}\left( \tau ,\bar{x}^{\prime
}\right) $, i.e., 
\begin{equation}
R^{ij}\left( \tau ,\bar{x},\bar{x}^{\prime }\right) =\overline{\mathtt{v}%
^{i}\left( \tau ,\bar{x}\right) \mathtt{v}^{j}\left( \tau ,\bar{x}^{\prime
}\right) }  \label{ia}
\end{equation}%
where supra-indexes indicate the Cartesian components of the turbulent
velocity. From eq. (\ref{g}), we can define a \textsl{stochastic velocity
operator} as%
\begin{equation}
\mathtt{v}_{\Phi }^{i}=\frac{T_{\Phi }^{\left\{ 0i\right\} }\left( \tau ,%
\bar{x}\right) }{\left\langle \rho +p\right\rangle \left( \tau \right) }
\label{ib}
\end{equation}%
and we assume that it is not relativistic. Observe that this does not mean
that the particles are non-relativistic, they indeed are at such high
energy. However their collective motion can be safely taken as
non-relativistic. A state of a quantum field is specified by its Hadamard
two point function, i.e., the vacuum expectation value of the anticommutator
of the field at different spacetime points. So from definition (\ref{ib}) we
can calculate $\left\langle 0\left\vert \left\{ T_{\Phi }^{\left\{
0i\right\} }\left( \tau ,\bar{x}\right) ,T_{\Phi }^{\left\{ 0j\right\}
}\left( \tau ^{\prime },\bar{x}^{\prime }\right) \right\} \right\vert
0\right\rangle $, and from it obtain the spatial two point correlation
function of the velocity field as 
\begin{equation}
R^{ij}\left( \tau ,\bar{x},\bar{x}^{\prime }\right) =\lim_{\tau ^{\prime
}\rightarrow \tau }\frac{\left\langle 0\left\vert \left\{ T_{\Phi }^{\left\{
0i\right\} }\left( \tau ,\bar{x}\right) ,T_{\Phi }^{\left\{ 0j\right\}
}\left( \tau ^{\prime },\bar{x}^{\prime }\right) \right\} \right\vert
0\right\rangle }{\left\langle \rho +p\right\rangle ^{2}\left( \tau \right) }
\label{ja}
\end{equation}%
using eqs. (\ref{jb}) and (\ref{jbb}), the Hadamard two point function reads%
\begin{align}
&\left\langle 0\left\vert
\left\{ T_{\Phi }^{\left\{ 0i\right\} }\left( \tau ,%
\bar{x}\right) ,T_{\Phi }^{\left\{ 0j\right\} }\left( \tau ^{\prime },\bar{x}%
^{\prime }\right) \right\} \right\vert 0\right\rangle \simeq \nonumber\\
&\simeq \frac{H^{8}}{%
32\pi ^{3}a^{6}\left( \tau \right) }\iint d\bar{\kappa}d\bar{\varpi}\left[
\varpi ^{i}\kappa ^{j}\frac{\partial }{\partial \tau }G_{\kappa }^{I+}\left(
\tau ,\tau ^{\prime }\right) \frac{\partial }{\partial \tau ^{\prime }}%
G_{\varpi }^{I+}\left( \tau ,\tau ^{\prime }\right) \right.  \notag \\
&+\kappa ^{i}\varpi ^{j}\frac{\partial }{\partial \tau }G_{\varpi
}^{I+}\left( \tau ,\tau ^{\prime }\right) \frac{\partial }{\partial \tau
^{\prime }}G_{\kappa }^{I+}\left( \tau ,\tau ^{\prime }\right) +\varpi
^{i}\varpi ^{j}\frac{\partial ^{2}G_{\kappa }^{I+}\left( \tau ,\tau ^{\prime
}\right) }{\partial \tau ^{\prime }\partial \tau }G_{\varpi }^{I+}\left(
\tau ,\tau ^{\prime }\right)  \label{corr-5} \\
&+\left. \kappa ^{i}\kappa ^{j}\frac{\partial ^{2}G_{\varpi }^{I+}\left(
\tau ,\tau ^{\prime }\right) }{\partial \tau ^{\prime }\partial \tau }%
G_{\kappa }^{I+}\left( \tau ,\tau ^{\prime }\right) \right] e^{i\left( \bar{%
\kappa}+\bar{\varpi}\right) .\bar{\xi}}+\left( \tau ,\bar{\xi}\right) \left.
\leftrightarrow \right. \left( \tau ^{\prime },-\bar{\xi}\right)  \notag
\end{align}
with $\bar{\xi}=\bar{x}-\bar{x}^{\prime }$, and where $G_{\kappa
}^{I+}\left( \tau ,\tau ^{\prime }\right) =\phi _{\kappa }^{I}\left( \tau
\right) \phi _{\kappa }^{I\ast }\left( \tau ^{\prime }\right) $ is the
positive frequency propagator.
Writing the scalar field modes during inflation, $\phi _{\kappa }^{I}\left(
\tau \right) $ in terms of the modes during reheating, $\phi _{\kappa
}^{R}\left( \tau \right) $ as $\phi _{\kappa }^{I}\left( \tau \right)
=\alpha _{\kappa }\phi _{\kappa }^{R}\left( \tau \right) +\beta _{\kappa
}\phi _{\kappa }^{R\ast }\left( \tau \right) $ with $\alpha _{\kappa }$ and $%
\beta _{\kappa }$ the Bogoliubov coefficients, we find the positive
frequency propagator 
\begin{align}
G_{\kappa }^{I+}\left( \tau ,\tau ^{\prime }\right) & =G_{\kappa }^{R+}\left(
\tau ,\tau ^{\prime }\right) +\alpha _{\kappa }\beta _{\kappa }^{\ast }\phi
_{\kappa }^{R}\left( \tau \right) \phi _{\kappa }^{R}\left( \tau ^{\prime
}\right) +\alpha _{\kappa }^{\ast }\beta _{\kappa }\phi _{\kappa }^{R\ast
}\left( \tau \right) \phi _{\kappa }^{R\ast }\left( \tau ^{\prime }\right)
\nonumber\\
&+\left\vert \beta _{\kappa }\right\vert ^{2}\left[ G_{\kappa }^{R+}\left(
\tau ,\tau ^{\prime }\right) +G_{\kappa }^{R-}\left( \tau ,\tau ^{\prime
}\right) \right]  \label{corr-6-a}
\end{align}
with $G_{\kappa }^{R-}\left( \tau ,\tau ^{\prime }\right) =\phi _{\kappa
}^{R\ast }\left( \tau \right) \phi _{\kappa }^{R}\left( \tau ^{\prime
}\right) $ the negative frequency propagator. When replacing eq. (\ref%
{corr-6-a}) in eq. (\ref{corr-5}) there appear three kernels, one with
vacuum contributions only, another with contributions both from the vacuum
and from the created particles, and a third one, built from contributions
from the created particles alone. This is the most important one. Details of
the calculations, as well as explanations of the approximations made along
the way, are given in Appendix C. Here we quote the main results and discuss
the physics involved. Replacing the propagators and their derivatives, and
neglecting rapidly decaying terms, we obtain%
\begin{align}
R^{ij}&\left( \tau ,\bar{x},\bar{x}+\bar{\xi}\right) \simeq \lim_{\tau
^{\prime }\rightarrow \tau }\frac{H^{8}}{32\pi ^{3}a^{6}\left( \tau \right) }%
\frac{1}{\left\langle \rho +p\right\rangle ^{2}}\iint d\bar{\kappa}d\bar{%
\varpi}~e^{i\left( \bar{\kappa}+\bar{\varpi}\right) .\bar{\xi}}\left\vert
\beta _{\kappa }\right\vert ^{2}\left\vert \beta _{\varpi }\right\vert ^{2}
\label{corr-9} \\
& \times \left\{ \varpi ^{i}\varpi ^{j}\left[ G_{\varpi }^{R+}\left( \tau
,\tau ^{\prime }\right) +G_{\varpi }^{R-}\left( \tau ,\tau ^{\prime }\right) %
\right] \left[ \frac{\partial ^{2}}{\partial \tau ^{\prime }\partial \tau }%
G_{\kappa }^{R+}\left( \tau ,\tau ^{\prime }\right) +\frac{\partial ^{2}}{%
\partial \tau ^{\prime }\partial \tau }G_{\kappa }^{R-}\left( \tau ,\tau
^{\prime }\right) \right] \right.  \notag \\
&+ \left. \kappa ^{i}\kappa ^{j}\left[ G_{\kappa }^{R+}\left( \tau ,\tau
^{\prime }\right) +G_{\kappa }^{R-}\left( \tau ,\tau ^{\prime }\right) %
\right] \left[ \frac{\partial ^{2}}{\partial \tau ^{\prime }\partial \tau }%
G_{\varpi }^{R+}\left( \tau ,\tau ^{\prime }\right) +\frac{\partial ^{2}}{%
\partial \tau ^{\prime }\partial \tau }G_{\varpi }^{R-}\left( \tau ,\tau
^{\prime }\right) \right] \right\}  \notag
\end{align}%
The quantity $\left\langle \rho +p\right\rangle $ is calculated in Appendix
B, and to obtain it we neglected its fluctuations. This means that we are
identifying it with its expectation value. The result is
\begin{equation}
\left\langle \rho +p\right\rangle \simeq \frac{H^{4}}{2\left( 2\pi \right)
^{1/2}a^{4}\left( \tau \right) }\frac{1}{\tau _{0}^{2}}\left[ \frac{3}{2}%
\left( \frac{H}{m}\right) ^{2}\tau _{0}^{2}+\left( \frac{9}{16}\right) ^{4}%
\frac{1}{24}\right] .  \label{rho-p}
\end{equation}%
Observe that it depends on two parameters, $m/H$ and $\tau _{0}$ which are
related to the contribution of super-horizon and sub-horizon modes
respectively. The velocity spectrum, $\Phi ^{ij}\left( \varsigma ,\tau
\right) $, is given by the Fourier transform of eq. (\ref{corr-9}), i.e.,%
\begin{equation}
\Phi ^{ij}\left( \varsigma ,\tau \right) =\frac{1}{\left( 2\pi \right) ^{3/2}%
}\int d^{3}\bar{\xi}R^{ij}\left( \tau ,\bar{r},\bar{r}+\bar{\xi}\right) e^{-i%
\bar{\varsigma}\cdot \bar{\xi}}  \label{corr-9-a}
\end{equation}%
As shown by Tomita et al \cite{tomita-70}, eddies larger than the horizon
are frozen in the plasma and decay with the expansion. We shall consider
only modes inside the particle horizon, i.e., modes that are in causal
connection, so $\varsigma \geq 1$. Further calculations are sketched in
Appendix C. The main contribution to the velocity spectrum is due to
sub-horizon modes, almost parallel to $\bar{\varsigma}$. After performing
the calculations eq. (\ref{corr-9-a}) reads%
\begin{align}
\Phi ^{ij}\left( \varsigma ,\tau \right) &\simeq \left( \frac{9}{16}\right)
^{4}\frac{3}{512}\frac{1}{\pi }\frac{a^{4}\left( \tau \right) }{H^{3}}\left( 
\frac{H}{m}\right) ^{4}\tau _{0}^{2}\left[ \frac{3}{2}\left( \frac{H}{m}%
\right) ^{2}\tau _{0}^{2}+\left( \frac{9}{16}\right) ^{4}\frac{1}{24}\right]
^{-2}\nonumber\\
&\times \left( 3\frac{\varsigma ^{i}\varsigma ^{j}}{\varsigma ^{11}}+\frac{%
\delta ^{ij}}{\varsigma ^{9}}\right)  \label{corr-9-b}
\end{align}%
The general form of $\Phi ^{ij}\left( \varsigma ,\tau \right) $ for
isotropic turbulence can be written as \cite{monin-yaglom}%
\begin{equation}
\Phi ^{ij}\left( \varsigma ,\tau \right) =\left[ \Phi _{LL}\left( \varsigma
,\tau \right) -\Phi _{NN}\left( \varsigma ,\tau \right) \right] \frac{%
\varsigma ^{i}\varsigma ^{j}}{\varsigma ^{2}}+\Phi _{NN}\left( \varsigma
,\tau \right) \delta ^{ij}  \label{corr-g}
\end{equation}%
where $\Phi _{LL}\left( \varsigma ,\tau \right) $ is the longitudinal part
of the spectrum and $\Phi _{NN}\left( \varsigma ,\tau \right) $ the normal
part. By direct comparison of eq. (\ref{corr-g}) with (\ref{corr-9-b}) we
have 
\begin{equation}
\Phi _{NN}\left( \varsigma ,\tau \right) \simeq \left( \frac{9}{16}\right)
^{4}\frac{3}{512}\frac{1}{\pi }\frac{a^{4}\left( \tau \right) }{H^{3}}\left( 
\frac{H}{m}\right) ^{4}\tau _{0}^{2}\left[ \frac{3}{2}\left( \frac{H}{m}%
\right) ^{2}\tau _{0}^{2}+\left( \frac{9}{16}\right) ^{4}\frac{1}{24}\right]
^{-2}\frac{1}{\varsigma ^{9}}  \label{phi-NN}
\end{equation}%
\begin{equation}
\Phi _{LL}\left( \varsigma ,\tau \right) \simeq \left( \frac{9}{16}\right)
^{4}\frac{3}{128}\frac{1}{\pi }\frac{a^{4}\left( \tau \right) }{H^{3}}\left( 
\frac{H}{m}\right) ^{4}\tau _{0}^{2}\left[ \frac{3}{2}\left( \frac{H}{m}%
\right) ^{2}\tau _{0}^{2}+\left( \frac{9}{16}\right) ^{4}\frac{1}{24}\right]
^{-2}\frac{1}{\varsigma ^{9}}  \label{phi-LL}
\end{equation}%
Observe that both functions are of similar amplitude. The energy spectrum is
defined as%
\begin{eqnarray}
E\left( \varsigma ,\tau \right) &=&\frac{H^{2}}{2}\int d\Omega \left( \bar{%
\varsigma}\right) \frac{\varsigma ^{2}}{a^{2}\left( \tau \right) }\Phi
^{ii}\left( \varsigma ,\tau \right)  \label{ene-spect-0} \\
&\simeq &\left( \frac{9}{16}\right) ^{4}\frac{9}{128}\frac{a^{2}\left( \tau
\right) }{H}\left( \frac{H}{m}\right) ^{4}\tau _{0}^{2}\left[ \frac{3}{2}%
\left( \frac{H}{m}\right) ^{2}\tau _{0}^{2}+\left( \frac{9}{16}\right) ^{4}%
\frac{1}{24}\right] ^{-2}\frac{1}{\varsigma ^{7}}  \notag
\end{eqnarray}%
where $d\Omega \left( \bar{\varsigma}\right) $ is the solid angle element,
and%
\begin{equation}
\Phi ^{ii}\left( \varsigma ,\tau \right) =2\Phi _{NN}\left( \varsigma ,\tau
\right) +\Phi _{LL}\left( \varsigma ,\tau \right)  \label{ene-spect-2}
\end{equation}%
The total energy per mass unit is then 
\begin{align}
E\left( \tau \right) & =\frac{1}{2}\left\langle \mathtt{v}^{2}\right\rangle
\simeq H\int_{1}^{\infty }\frac{d\varsigma }{a\left( \tau \right) }E\left(
\varsigma ,\tau \right)  \label{tot-ener-1} \\
& \simeq \left( \frac{9}{16}\right) ^{4}\frac{3}{256}a\left( \tau \right)
\left( \frac{H}{m}\right) ^{4}\tau _{0}^{2}\left[ \frac{3}{2}\left( \frac{H}{%
m}\right) ^{2}\tau _{0}^{2}+\left( \frac{9}{16}\right) ^{4}\frac{1}{24}%
\right] ^{-2}  \notag
\end{align}%
We may define the total energy associated to a scale $\kappa \geq 1$ as
given by the contribution of all eddies smaller than $\kappa ^{-1}$, i.e.,%
\begin{equation}
E_{\kappa }\left( \tau \right) \equiv \frac{1}{2}\left\langle \mathtt{v}%
_{\kappa }^{2}\right\rangle =H\int_{\kappa }^{\infty }\frac{d\varsigma }{%
a\left( \tau \right) }E\left( \varsigma ,\tau \right)  \label{energia-k}
\end{equation}%
and so%
\begin{equation}
E_{\kappa }\left( \tau \right) \simeq \left( \frac{9}{16}\right) ^{4}\frac{3%
}{256}a\left( \tau \right) \left( \frac{H}{m}\right) ^{4}\tau _{0}^{2}\left[ 
\frac{3}{2}\left( \frac{H}{m}\right) ^{2}\tau _{0}^{2}+\left( \frac{9}{16}%
\right) ^{4}\frac{1}{24}\right] ^{-2}\frac{1}{\kappa ^{6}}
\label{energia-k-2}
\end{equation}

\subsection{Kinetic helicity two point correlation function}

A sufficient condition to sustain large scale dynamo action is that the
turbulence be helical \cite{moffatt,dan-pab-1,mininni-07}. As stated in
Sect. II, it is defined as the volume average of the scalar product of the
vorticity and the velocity \cite{lesieur}, i.e.,%
\begin{equation}
\mathcal{H}_{c}=\overline{\mathtt{w}^{i}\mathtt{v}^{i}}\equiv \frac{1}{Vol}%
\int_{Vol}d\left( vol\right) ~\mathtt{w}^{i}\mathtt{v}^{i}
\label{kin-hel-def}
\end{equation}%
with $\mathtt{w}^{i}=\epsilon ^{ijk}\partial _{j}\mathtt{v}_{k}$ the
vorticity of the velocity field. A non-null value of this quantity indicates
that the flow lacks of mirror symmetry. Due to conservation of angular
momentum in the particle creation process, the expectation value of the
kinetic helicity must vanish. However the r.m.s. value, or variance, can be
different from zero, and this is what we show in this subsection.

\begin{figure}[ht]
\includegraphics{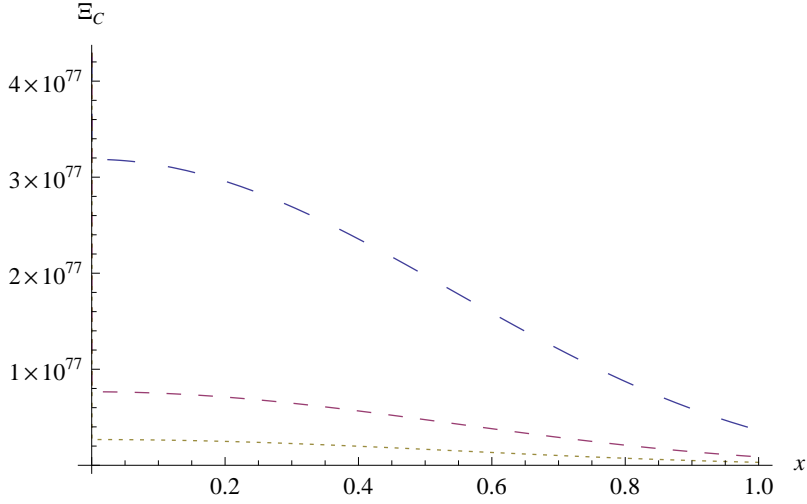}\vspace{0.5cm}
\caption{$16\pi^6 \left[a\left( \tau \right)/H\right]^2 \Xi _{c}^{\Phi }\left( \tau ,%
\bar{\xi}\right)$ as a
function of $\protect\xi$, for fixed $m/H=10^{-6}$ and $\protect\tau_0 =
0.7\times 10^{-9}$ (large dashing), $\protect\tau_0 = 10^{-9}$ (medium
dashing) and $\protect\tau_0 = 1.3\times 10^{-9}$ (tiny dashing)}
{\label{corr-hc-1}}
\end{figure}

\begin{figure}[ht]
\includegraphics{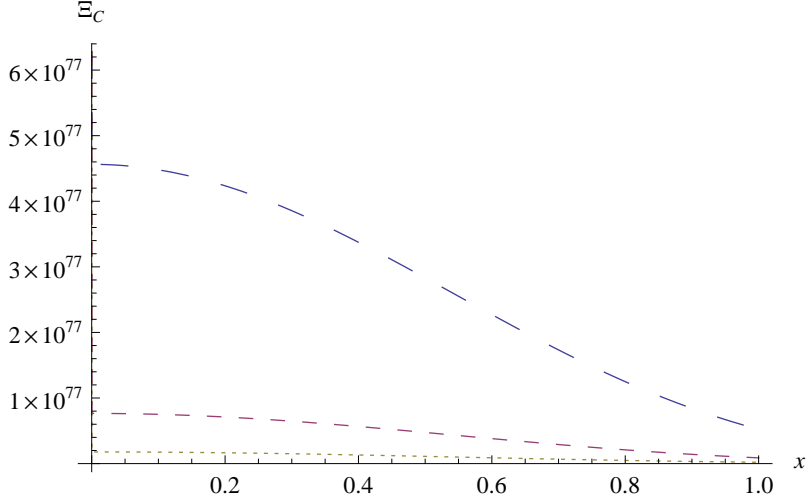} \vspace{0.5cm}
\caption{$16\pi^6 \left[a\left( \tau \right)/H\right]^2 \Xi _{c}^{\Phi }\left( \tau ,%
\bar{\xi}\right)$ as a
function of $\protect\xi$, for fixed $\protect\tau_0=10^{-9}$ and $m/H =
0.8\times 10^{-6}$ (large dashing), $m/H = 10^{-6}$ (medium dashing) and $%
m/H = 1.2\times 10^{-9}$ (tiny dashing)}
\label{corr-hc-2}
\end{figure}

From equation (\ref{ib}) we can write a \textsl{vorticity operator} as 
\begin{equation}
\mathtt{w}^{i}=\epsilon _{ijk}\partial _{j}\left[ \frac{T_{\Phi }^{\left\{
0k\right\} }}{\left\langle \rho +p\right\rangle }\right]  \label{kin-hel-2}
\end{equation}%
and define a \textsl{kinetic helicity operator} as%
\begin{equation}
H_{c}^{\Phi }=\frac{\epsilon _{ijk}T_{\Phi }^{\left\{ 0i\right\} }\partial
_{j}T_{\Phi }^{\left\{ 0k\right\} }}{4\left\langle \rho +p\right\rangle ^{2}}
\label{kin-hel-3}
\end{equation}%
which in terms of the fields reads%
\begin{align}
H_{c}^{\Phi }=\frac{H^{9}}{8\left\langle \rho +p\right\rangle ^{2}}\epsilon
^{ijk}& \left[ \left( \partial _{i}\Phi \right) \dot{\Phi}^{\dagger }\left(
\partial _{j}\dot{\Phi}\right) \left( \partial _{k}\Phi ^{\dagger }\right)
+\left( \partial _{j}\dot{\Phi}\right) \left( \partial _{k}\Phi ^{\dagger
}\right) \left( \partial _{i}\Phi \right) \dot{\Phi}^{\dagger }\right.
\label{kin-hel-4} \\
& +\left. \dot{\Phi}\left( \partial _{i}\Phi ^{\dagger }\right) \left(
\partial _{k}\Phi \right) \left( \partial _{j}\dot{\Phi}^{\dagger }\right)
+\left( \partial _{k}\Phi \right) \left( \partial _{j}\dot{\Phi}^{\dagger
}\right) \dot{\Phi}\left( \partial _{i}\Phi ^{\dagger }\right) \right] 
\notag
\end{align}%
Observe that in principle it does not vanish identically because there are
two fields involved in its expression. The r.m.s. value of $H_{c}^{\Phi }$
is again given by the vacuum expectation value of the Hadamard two point
function, i.e., $\left\langle 0\left\vert \left\{ H_{c}^{\Phi }\left( \bar{r}%
,\tau \right) ,H_{c}^{\Phi }\left( \bar{r}^{\prime },\tau ^{\prime }\right)
\right\} \right\vert 0\right\rangle $ from where we obtain a spatial two
point function as 
\begin{equation}
\Xi _{c}^{\Phi }\left( \tau ,\bar{x},\bar{x}^{\prime }\right) =\lim_{\tau
^{\prime }\rightarrow \tau }\left\langle 0\left\vert \left\{ H_{c}^{\Phi
}\left( \bar{x},\tau \right) ,H_{c}^{\Phi }\left( \bar{x}^{\prime },\tau
^{\prime }\right) \right\} \right\vert 0\right\rangle  \label{kin-hel-5}
\end{equation}%
The calculations are rather long but straightforward; details are given in
Appendix D. We quote here the main results. When replacing the fields we
obtain, as in the case of the velocity correlation $R_{ij}$, several
kernels: one with vacuum contributions only, another with mixed
contributions from vacuum and created particles, and a third with
contributions from the created particles only. Terms containing $\left\vert
\beta _{k}\right\vert ^{2}$ give the main contribution, because, as was the
case for $R_{ij}$, terms with $\alpha _{\kappa }\alpha _{\varpi }^{\ast
}\ldots $, etc. oscillate, and will give negligible contributions when
integrated.  In Fig. (\ref{corr-hc-1}) we show the dependence of 
$16\pi^6 \left[a\left( \tau \right)/H\right]^2 \Xi _{c}^{\Phi }\left( \tau ,%
\bar{\xi}\right)$ on $\xi$ for fixed $m/H = 10^{-6}$ and three values of $%
\tau _{0}$ and in Fig. (\ref{corr-hc-2})  the dependence on $\xi$ for fixed $%
\tau _{0} = 10^{-9}$ and three values of $m/H$.

\begin{figure}[ht]
\includegraphics{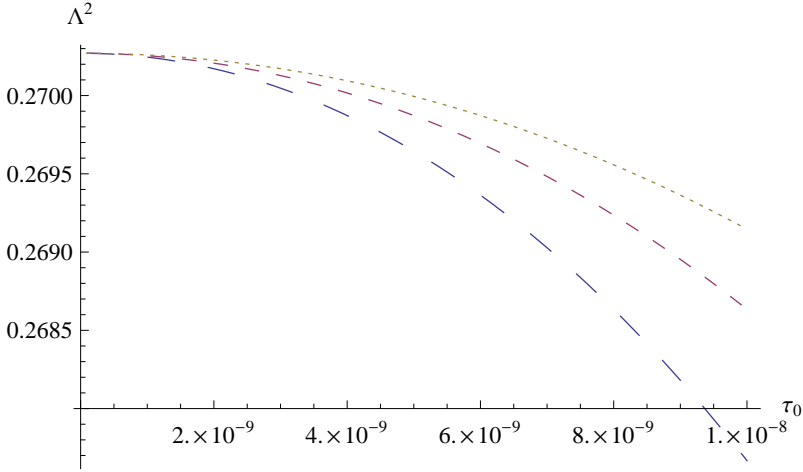} \vspace{0.5cm}
\caption{$\Lambda^2$ as a function of $\protect\tau_0$, and for $m/H =
0.8\times 10^{-6}$ (large dashing), $m/H = 10^{-8}$ (medium dashing) and $%
m/H = 1.2\times 10^{-9}$ (tiny dashing)}
\label{lambda-1}
\end{figure}

\begin{figure}[ht]
\includegraphics{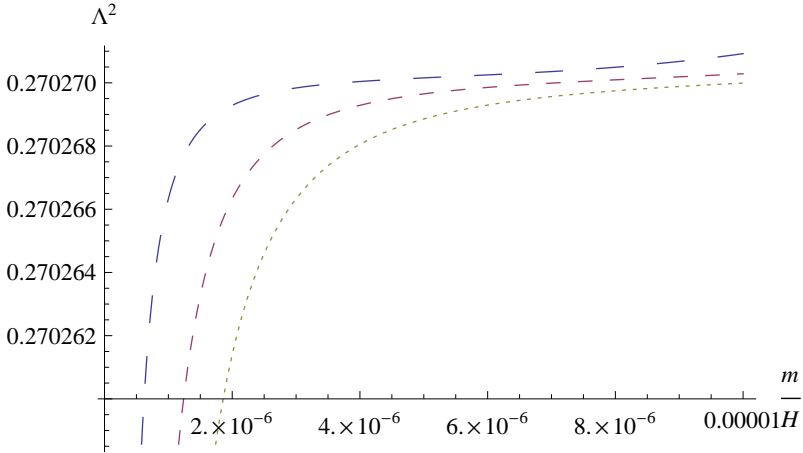} \vspace{0.5cm}
\caption{$\Lambda^2$ as a function of $m/H$, and for $\protect\tau_0 =
0.7\times 10^{-9}$ (large dashing), $\protect\tau_0 = 10^{-9}$ (medium
dashing) and $\protect\tau_0 = 1.3\times 10^{-9}$ (tiny dashing)}
\label{lambda-2}
\end{figure}

We can estimate the (dimensionless) coherence length of the kinetic helicity
as 
\begin{equation}
\Lambda ^{2}\left( \frac{m}{H},\tau _{0}\right) \equiv -\left. \frac{\Xi
_{c}^{\Phi }\left( \tau ,\bar{\xi}\right) }{\partial ^{2}\Xi _{c}^{\Phi
}\left( \tau ,\bar{\xi}\right) /\partial \xi ^{2}}\right\vert _{\xi =0}
\label{kin-hel-6b}
\end{equation}%
In Fig. (\ref{lambda-1}) we show $\Lambda ^{2}$ as a function of $m/H$ for $%
\tau _{0}$ fixed, and in Fig. and (\ref{lambda-2}) the converse. From both
figures we see that, for the chosen values of the parameters, $\Lambda \sim
1 $, i.e. it is of the order of the particle horizon's scale.

The r.m.s. value of the kinetic helicity is obtained by volume averaging
expression (\ref{kin-hel-5}) over $x$ and $x^{\prime }$, i.e.%
\begin{equation}
\Sigma_{\mathcal{H}_c}^2=\frac{1}{Vol\left( x\right) }%
\int_{Vol\left( x\right) }d\left[ vol\left( x\right) \right] \frac{1}{%
Vol\left( x^{\prime }\right) }\int_{Vol\left( x^{\prime }\right) }d\left[
vol\left( x^{\prime }\right) \right] ~\Xi _{c}^{\Phi }\left( \tau ,\bar{x},%
\bar{x}^{\prime }\right)   \label{kin-hel-6}
\end{equation}%
Kinetic helicity is a global quantity, that
can depend at most on the dimensionless characteristic scale $\mathcal{L}<1$
of the integration volume. To evaluate the integrals in eq. (\ref{kin-hel-6}%
) we can proceed as follows. As we are considering scales $\lesssim 1$ we
can develop eq. (\ref{kin-hel-5}) in Taylor series to second order in $\bar{x%
}-\bar{x}^{\prime }$\footnote{%
It can be seen from eq. (\ref{ap4-8}) that this is the next to leading order.%
}, then using eq. (\ref{rho-p}) we have 
\begin{align}
\Sigma_{\mathcal{H}_c}\left( \tau ,\mathcal{L}\right) &\sim \frac{8H\tau _{0}^{4}}{%
\pi a\left( \tau \right) }\left[ \frac{3}{2}\left( \frac{H}{m}\right)
^{2}\tau _{0}^{2}+\left( \frac{9}{16}\right) ^{4}\frac{1}{24}\right]
^{-2}\nonumber\\
&\times A^{1/2}\left( \frac{m}{H},\tau _{0}\right) \left[ 1-\frac{1}{240}\frac{%
\mathcal{L}^{2}}{\Lambda ^{2}\left( m/H,\tau _{0}\right) }\right] ^{1/2}
\label{kin-hel-7}
\end{align}%
where $\mathcal{L}=HL$, and with%
\begin{align}
A\left( \frac{m}{H},\tau _{0}\right) & \simeq  -\frac{2187}{2097152}\frac{1}{%
\tau _{0}^{2}}\left[ \frac{3}{8}\left( \frac{H}{m}\right) ^{4}+\frac{1}{2}%
\left( \frac{H}{m}\right) ^{2}+\frac{9477}{29360128}\frac{1}{\tau _{0}^{2}}%
\right] \nonumber\\
& \times \left\{ \left[ \frac{1}{12}\frac{H}{m}-\frac{2187}{5242880}\frac{1}{%
\tau _{0}^{2}}\right] ^{2}
- \left[ \frac{3}{8}\left( \frac{H}{m}\right) ^{2}+\frac{2187}{%
4194304}\frac{1}{\tau _{0}^{2}}\right] \right.\nonumber\\
& \times \left.\left[ \frac{3}{8}\left( \frac{H}{m}%
\right) ^{4}+\frac{1}{2}\left( \frac{H}{m}\right) ^{2}+\frac{9477}{29360128}%
\frac{1}{\tau _{0}^{2}}\right] \right\}   \label{kin-hel-8}
\end{align}%
the zeroth order in the Taylor expansion of $\Xi _{c}^{\Phi }\left( \tau ,%
\bar{\xi}\right) $.

\section{Characterizing the turbulence: viscosity, Reynolds number and
characteristic decay and correlation times}

Turbulence sets in at the inflation-reheating transition, and afterward, as
there is no stirring forces acting on the flow, it decays. Since the fluid
is made of particle - antiparticle pairs, annihilation also occurs. We have
two competing processes, whose characteristic times must be compared in
order to decide which one dominates: the decay time of turbulence and the
time of particle anti-particle annihilation.

\begin{figure}[ht]
\includegraphics{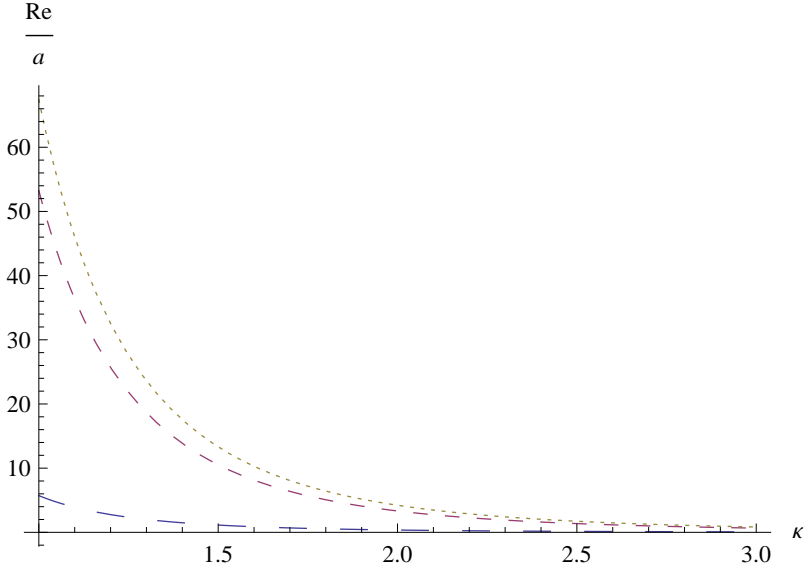} \vspace{0.5cm}
\caption{$Re\left(\protect\kappa ,\protect\tau\right) /a^{1/2}\left(\protect\tau%
\right)$ as a function of $\protect\kappa$, for fixed $m/H=10^{-5}$ and $%
\protect\tau_0 = 10^{-7}$ (large dashing), $10^{-8}$ (medium dashing) and $%
10^{-9}$ (tiny dashing). $Re$ increases as $\protect\tau_0$ diminishes}
\label{Re-k-1}
\end{figure}

\begin{figure}[ht]
\includegraphics{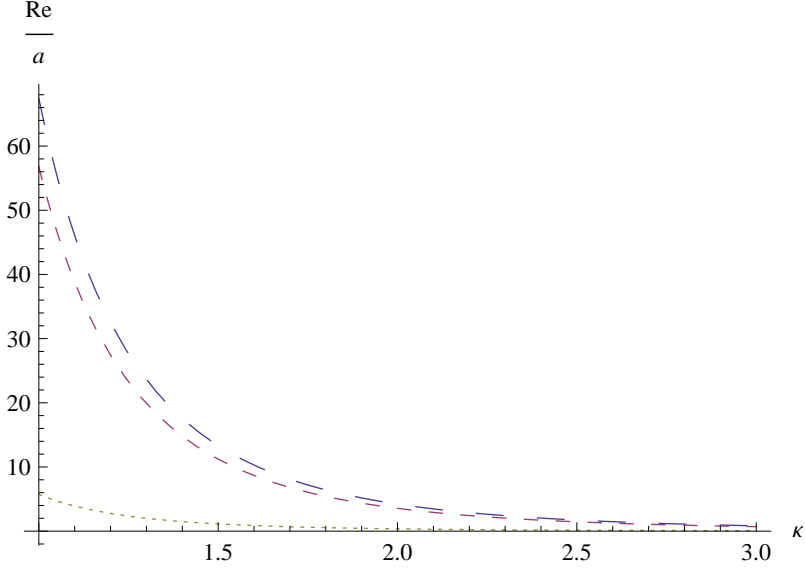} \vspace{0.5cm}
\caption{$Re\left(\protect\kappa ,\protect\tau\right) /a^{1/2}\left(\protect\tau%
\right)$ as a function of $\protect\kappa$, for fixed $\protect\tau%
_0=10^{-9} $ and $m/H = 10^{-5}$ (large dashing), $10^{-6}$ (medium dashing)
and $10^{-7}$ (tiny dashing). $Re$ diminishes as $m/H$ diminishes}
\label{Re-k-2f}
\end{figure}

\begin{figure}[ht]
\includegraphics{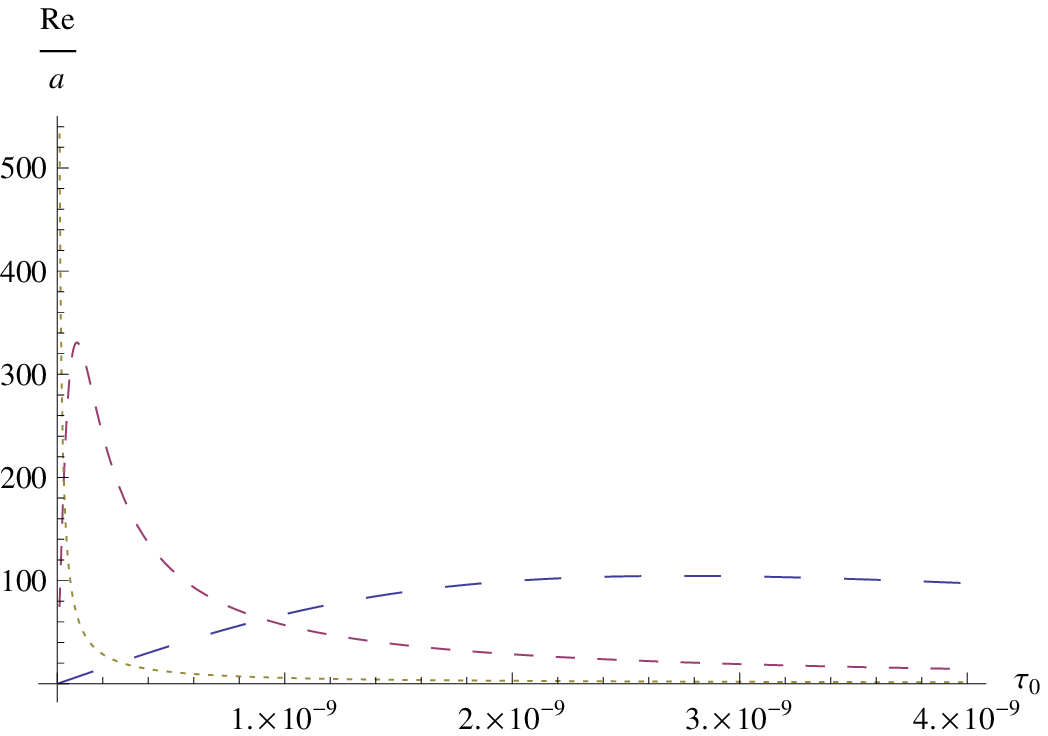} \vspace{0.5cm}
\caption{$Re\left(\protect\kappa =1 ,\protect\tau\right) /a^{1/2}\left(\protect\tau%
\right)$ as a function of $\protect\tau_0$, and for $m/H = 10^{-5}$ (large
dashing), $10^{-6}$ (medium dashing) and $10^{-7}$ (tiny dashing). Observe
that $Re$ peaks at a certain value of $\protect\tau_0$. The peak is higher
and occurs at smaller values of $\protect\tau_0$ as $m/H$ diminishes. }
\label{Re-t}
\end{figure}

\begin{figure}[ht]
\includegraphics{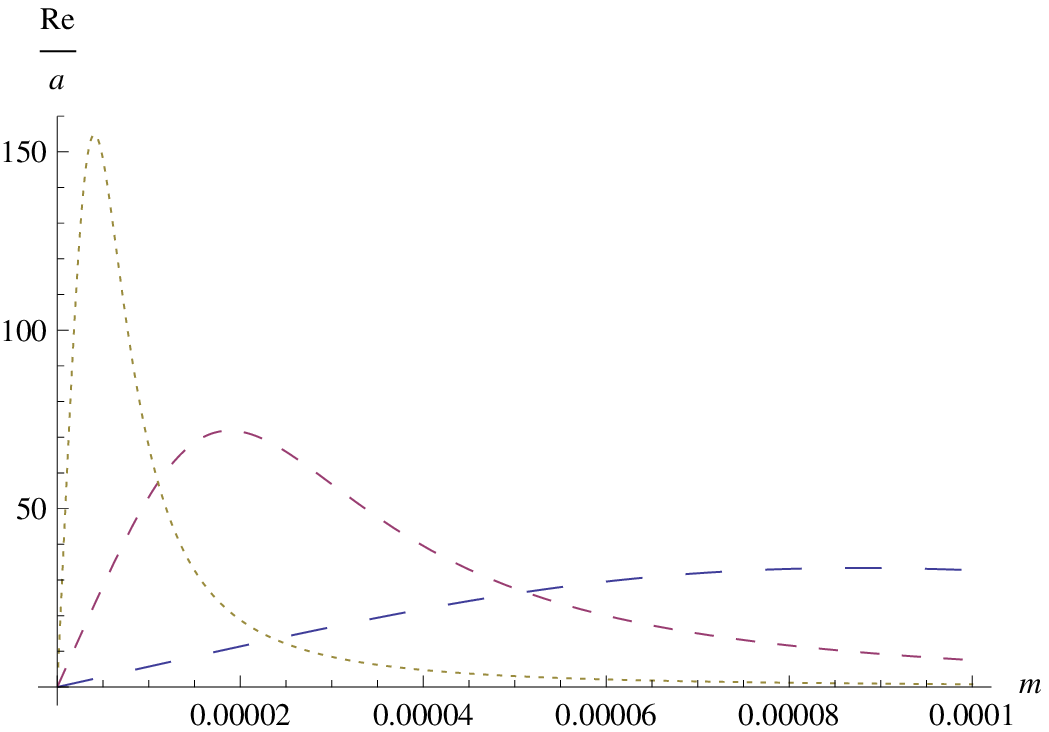} \vspace{0.5cm}
\caption{$Re\left(\protect\kappa =1 ,\protect\tau\right) /a^{1/2}\left(\protect\tau%
\right)$ as a function of $m/H$, and for $\tau_0 = 10^{-7}$ (large dashing), $%
10^{-8}$ (medium dashing) and $10^{-9}$ (tiny dashing). Observe that $Re$
again peaks at a certain value of $m/H$. The peak is higher and occurs at
smaller values of $m/H$ as $\protect\tau_0$ diminishes. }
\label{Re-m}
\end{figure}

To properly characterize the turbulence, we must first calculate the
Reynolds number of the flow. This number is defined as $Re=ul/\upsilon $,
with $u$ and $l$ characteristics velocity and scale respectively and $%
\upsilon $ the kinematic viscosity. As it may happen that turbulence is not
fully developed at all scales, we must calculate this number for each scale.
The scale dependent Reynolds number can be written as 
\begin{equation}
Re\left( \kappa \right) =\frac{k^{-1}\mathtt{v}_{\kappa }}{\upsilon }
\label{Re-1}
\end{equation}%
where $\upsilon =\eta /\left\langle \rho +p\right\rangle $ is the dimensionless 
ratio of
the fluid shear viscosity to the energy density $k^{-1}=a\left( \tau \right)
H^{-1}\kappa ^{-1}$ is the scale of interest and $\mathtt{v}_{\kappa }\sim 
\sqrt{2E_{\kappa }\left( \tau \right) }$ the estimate of the velocity at the
corresponding scale. To estimate $\eta $ we follow the work of Son and
collaborators \cite{son-1}, and consider it proportional to the entropy
density, i.e., $\eta /s=1/4\pi $. We take $s$ as proportional to the
quasiparticle number density $n$, i.e., $s\sim n\left( \tau \right) $, with $%
n\left( \tau \right) $ 
\begin{equation}
n\left( \tau \right) \simeq \frac{H^{3}}{a^{3}\left( \tau \right) }\frac{1}{%
\tau _{0}^{2}}\left[ \frac{3}{2}\frac{H^{3}}{m^{3}}\tau _{0}^{2}+\left( 
\frac{9}{16}\right) ^{4}\frac{\pi }{28}\right]  \label{2c}
\end{equation}%
(see Appendix B). Using eq. (\ref{rho-p}) we have that%
\begin{equation}
\upsilon \left( \tau \right) \simeq \frac{1}{\left( 2\pi \right) ^{1/2}}%
\frac{a\left( \tau \right) }{H}\frac{\left[ \left( 3/2\right) \left(
H/m\right) ^{3}\tau _{0}^{2}+\left( 9/16\right) ^{4}\pi /28\right] }{\left[
\left( 3/2\right) \left( H/m\right) ^{2}\tau _{0}^{2}+\left( 9/16\right)
^{4}/24\right] }  \label{2d}
\end{equation}%
Replacing everything into eq. (\ref{Re-1}) we obtain%
\begin{equation}
Re\left( \kappa ,\tau \right) \simeq \left( 2\pi \right) ^{1/2}\left( \frac{9%
}{16}\right) ^{2}\sqrt{\frac{3}{256}}\frac{a^{1/2}\left( \tau \right) \left(
H/m\right) ^{2}\tau _{0}}{\left[ \left( 3/2\right) \left( H/m\right)
^{3}\tau _{0}^{2}+\left( 9/16\right) ^{4}\pi /28\right] }\frac{1}{\kappa ^{4}%
}  \label{Re-k}
\end{equation}%
We see that, for $\kappa \sim 1$, we can have $Re\gg 1$ if%
\begin{equation}
\left( 2\pi \right) ^{1/2}\frac{2}{3}\left( \frac{9}{16}\right) ^{2}\sqrt{%
\frac{3}{256}}\gg \frac{H}{m}\tau _{0}  \label{Re-k-2}
\end{equation}%
i.e., if the transition between inflation and reheating is very fast. In
Fig. (\ref{Re-k-1}) we plot $Re\left( \kappa ,\tau \right) /a^{1/2}\left(
\tau \right) $ as a function of $\kappa $ for fixed $m/H$ and three values
of $\tau _{0}$, it is seen that $Re$ increases as the duration of the
transition inflation/reheating shortens. In Fig. (\ref{Re-k-2f}) we plot the
same as in the first figure, but with fixed $\tau_0 = 10^{-9}$ and $m/H =
10^{-5}$, $10^{-6}$ and $10^{-7}$, and we observe that $Re$ diminishes with
decreasing $m/H$. In both figures $Re$ is a decreasing function of $\kappa$,
hence only the modes near the horizon can be considered as turbulent. In
Fig. (\ref{Re-t}) we plot $Re\left(\kappa =
1,\tau\right)/a^{1/2}\left(\tau\right)$ as a function of $\tau_0$ and for $%
m/H = 10^{-5}$, $10^{-6}$ and $10^{-7}$. In this case $Re$ has a peak at
certain value of $\tau_0$, and this peak is higher and occurs at shorter
values of $\tau_0$ as $m/H$ diminishes. As $\tau_0$ grows $Re$ decreases
monotonically. Finally, in Fig. (\ref{Re-m}) we show $Re/a^{1/2}\left(\tau%
\right)$ as a function of $m/H$, for $\kappa = 1$ and $\tau_0= 10^{-7}$, $%
10^{-8}$ e $10^{-9}$. Again $Re$ peaks at certain values of $m/H$ and the
peak is higher and occurs at smaller values of $m/H$ as $\tau_0$ diminishes.

The decay time of each turbulent mode is given by $t_{d}\left( k\right)
=1/\upsilon k^{2}=a^{2}\left( \tau \right) H^{-2}/\upsilon \kappa ^{2}$.
Using eq. (\ref{2d}) we can write%
\begin{equation}
t_{d}\left( \kappa \right) \simeq \frac{\left( 2\pi \right) ^{1/2}a\left(
\tau \right) }{H}\frac{\left[ \left( 3/2\right) \left( H/m\right) ^{2}\tau
_{0}^{2}+\left( 9/16\right) ^{4}/24\right] }{\left[ \left( 3/2\right) \left(
H/m\right) ^{3}\tau _{0}^{2}+\left( 9/16\right) ^{4}\pi /28\right] }\frac{1}{%
\kappa ^{2}}  \label{t-dec}
\end{equation}

To estimate the pair annihilation time of each mode, we consider that the
particles are relativistic, according to the result obtained for their
energy density. On a dimensional basis, this time can be estimated as%
\begin{equation}
t_{a}\left( \kappa \right) \sim \frac{1}{n\left( \tau \right) \sigma
_{\kappa }u_{r}}  \label{2f}
\end{equation}%
with $n\left( \tau \right) $ the particle density, $\sigma _{\kappa }$ the
annihilation cross-section and $u_{r}$ the relative velocity between species
which we take $u_{r}\sim 1$. We make a crude estimation of the cross
section, as being the same as for $e^{+}e^{-}$ annihilation \cite{Itzykson}
for $\gamma \gg 1$. i.e., $\sigma \simeq \pi r_{0}^{2}/\gamma $, with, $%
r_{0}=\alpha /m$, $\alpha =1/137$ the fine-structure constant, $\gamma
=\varepsilon /m$ with $m$ the particles mass and $\varepsilon \sim H\Omega
_{\kappa }=H\left[ \kappa ^{2}/a^{2}\left( \tau \right) +\left( m/H\right)
^{2}\right] ^{1/2}$ the maximum energy of each mode. We then have%
\begin{equation}
\sigma _{\kappa }\simeq \pi \frac{\alpha ^{2}}{mH}\frac{a\left( \tau \right) 
}{\sqrt{\kappa ^{2}+a^{2}\left( \tau \right) \left( m/H\right) ^{2}}}
\label{sigma-k}
\end{equation}%
replacing in eq. (\ref{2f}) we have%
\begin{equation}
t_{a}\left( \kappa \right) \sim \frac{a^{2}\left( \tau \right) }{\pi \alpha
^{2}}\frac{m\tau _{0}^{2}}{H^{2}}\frac{\sqrt{\kappa ^{2}+a^{2}\left( \tau
\right) \left( m/H\right) ^{2}}}{\left[ \left( 3/2\right) \left( H/m\right)
^{3}\tau _{0}^{2}+\left( 9/16\right) ^{4}/28\right] }  \label{t-aniq}
\end{equation}%
Comparing both times we have%
\begin{equation}
\frac{t_{d}\left( \kappa \right) }{t_{a}\left( \kappa \right) }\simeq \frac{%
\left( 2\pi ^{3}\right) ^{1/2}\alpha ^{2}}{a\left( \tau \right) }\frac{H}{%
m\tau _{0}^{2}}\frac{\left[ \left( 3/2\right) \left( H/m\right) ^{2}\tau
_{0}^{2}+\left( 9/16\right) ^{4}/24\right] }{\kappa ^{2}\sqrt{\kappa
^{2}+a^{2}\left( \tau \right) \left( m/H\right) ^{2}}}  \label{2h}
\end{equation}%
as $H/m\gg 1$ and $\tau _{0}^{2}\ll 1$ we have that $t_{d}\left( \kappa
\right) /t_{a}\left( \kappa \right) \gg 1$, i.e., annihilation occurs before
the end of turbulence.

It can be seen that $t_a$ is also much shorter than other time scales
pertaining to the flow, such as the ratio between the radius of the largest
turbulent eddy (i.e., the horizon as it is there where $Re\left(\kappa%
\right) $ takes its largest value) to the velocity associated to that scale.
Therefore in what follows we take $t_{corr}\equiv t_a$.

\section{Magnetic field amplification due to dynamo action}

According to eq. (\ref{mft-7}) we must now evaluate the amplification
exponent, $\Sigma{\mathcal{H}_{c}}^2\left( \tau ,\mathcal{L}\right) t\left( \kappa
\right) /\mathtt{L}a\left( \tau \right) $, where $\mathtt{L}a\left( \tau
\right) $ is the physical coherence scale of the magnetic field and $%
\mathcal{L}$ is the (dimensionless) coherent scale of the kinetic helicity.
From the discussion in Section III, the shortest time is the annihilation
time, which for scales such that $\kappa \sim 1/\mathcal{L}>m/H$ reads 
\begin{equation}
t_{a}\left( \mathcal{L}\right) \sim \frac{a^{2}\left( \tau \right) }{\pi
\alpha ^{2}}\frac{m\tau _{0}^{2}}{H^{2}}\frac{1}{\left[ \left( 3/2\right)
\left( H/m\right) ^{3}\tau _{0}^{2}+\left( 9/16\right) ^{4}/28\right] }\frac{%
1}{\mathcal{L}}  \label{am-1}
\end{equation}%
Writing $\mathtt{L}=H^{-1}\mathfrak{L}$ the amplification exponent reads 
\begin{eqnarray}
\frac{\Sigma_{\mathcal{H}_{c}}\left( \tau ,\mathcal{L}\right) t_{a}\left( 
\mathcal{L}\right) }{\mathtt{L}a\left( \tau \right) } &\sim &\frac{8}{\pi
^{3}\alpha ^{4}}a^{3}\left( \tau \right) \frac{m^{2}\tau _{0}^{8}}{H^{2}}%
\frac{1}{\mathfrak{L}\mathcal{L}^{2}}\frac{1}{\left[ \left( 3/2\right)
\left( H/m\right) ^{2}\tau _{0}^{2}+\left( 9/16\right) ^{4}1/24\right] ^{2}}
\notag \\
&&\times \frac{1}{\left[ \left( 3/2\right) \left( H/m\right) ^{3} \tau
_{0}^{2}+\left( 9/16\right) ^{4}/28\right] ^{2}}  \notag \\
&&\times A^{1/2}\left( \frac{m}{H},\tau _{0}\right) \left[ 1-\frac{1}{240}%
\frac{\mathcal{L}^{2}}{\Lambda ^{2}\left( m/H,\tau _{0}\right) }\right]^{1/2}
  \label{am-2}
\end{eqnarray}%
which by simple inspection is seen to be very small.

\section{Discussion and Conclusions}

In this paper we have studied the possibility of turbulent dynamo action
during reheating. We considered the presence of a charged scalar field
minimally coupled to gravity. This field is in its invariant vacuum state
during inflation. When the transition to reheating takes place the vacuum
state turns into a many-particle state. For sub-horizon modes of the field,
the number of created modes depends on the details of the transition.
Therefore during reheating, besides the decay products of the inflaton we
also have a plasma of scalar particles which is at rest in the comoving
frame. We characterize the fluctuating velocities of this plasma giving
their spatial two point correlation function and the kinetic energy
associated to each Fourier mode of the stochastic velocity field, eq. (\ref%
{energia-k}). We evaluate the Reynolds number associated to each mode, $%
Re\left( \kappa \right) $, which turns out to depend on the physical
parameters of the problem, namely $m$, $H$ and the duration $\tau _{0}$ of
the transition from inflation to reheating. If $\tau _{0}$ is small enough,
then there is a range of $\kappa $ for which $Re\left( \kappa \right) >1$
and the flow can be considered as (mildly) turbulent. As there are no
stirring forces, the turbulence we refer to decays, each mode doing so in a
characteristic time $t_{d}\left( \kappa \right) $ given by eq. (\ref{t-dec}%
). Besides as the plasma is a particle anti-particle one, each mode of the
scalar field (not to be confused with modes of the stochastic velocity
field) annihilates in a characteristic time $t_{a}\left( \kappa \right) $
given by eq. (\ref{t-aniq}). When comparing both times we find that
annihilation dominates over decay, eq. (\ref{2h}) and hence for practical
purposes we can consider the turbulence as steady.

The sufficient condition to have a large scale kinematic dynamo is the flow
to be endowed with kinetic helicity \cite{moffatt}. The non-trivial result
of this paper is that the scalar plasma does possess a non null rms kinetic
helicity,eq. (\ref{kin-hel-7}) From Figs. (\ref{lambda-1}) and (\ref%
{lambda-2}) we see that, for the parameters for which $Re\left( \kappa
\right) >1$, the characteristic scale of the kinetic helicity is of the
order of the particle horizon, thus allowing for kinematic dynamo action.

The existence of an rms helicity is due to the presence of the two scalar
fields, $\Phi $ and $\Phi ^{\dagger }$, as is evident from eq. (\ref%
{kin-hel-4}). Moreover, though the helicity may have either sign, in the
average the amplification effect dominates. From the simplest model of
kinematic dynamo, eq. (\ref{mft-6}), we compute the amplification factor of
an initial seed field, eq. (\ref{am-2}), and find that for the physical
parameters of the scenario considered, it is very small. In spite of this
result, we believe our work shows the need for exploring the impact of
nonlinear effects in the early universe. These effects offer the most
natural answer to the riddle of the survival of the primordial magnetic
field until the epoch of structure formation, in spite of the $1/a^{2}$
damping induced by the Hubble expansion.

\begin{acknowledgments}
We thank P. Mininni and D. Gomez for clarifying comments and discussions. E.
C. acknowledges support from University of Buenos Aires, CONICET and ANPCyT.
A.K. thanks CNPq for financial support through project CNPq/471254/2008-8.
\end{acknowledgments}

\appendix

\section{Bogoliubov coefficients}

We assume that during the reheating period the scale factor of the Universe
scales as $t^{2/3}$ \cite{kolb-turner,starob-80}, while for inflation we
consider a spatially flat de Sitter universe. For large wavenumbers the
Bogoliubov coefficients are sensitive to the details of the transition,
while for small wavenumbers the coefficients can be found assuming an
instantaneous transition. This dependence on the transition details for
subhorizon modes was also recently analyzed by Zaballa and Sasaki \cite%
{zab-sas-09} in the context of creation of metric perturbations at the end
of inflation.

The Klein Gordon equation for a free field in a FRW Universe is%
\begin{equation}
\ddot{\psi}_{\kappa }+\left[ \frac{\kappa ^{2}}{a^{2}}+\left( \frac{m}{H}%
\right) ^{2}-\frac{3}{2}\left( \frac{\ddot{a}}{a}+\frac{\dot{a}^{2}}{2a^{2}}%
\right) \right] \psi _{\kappa }=0  \label{bog-1-a}
\end{equation}%
It is seen that%
\begin{equation}
\frac{3}{2}\left( \frac{\ddot{a}}{a}+\frac{\dot{a}^{2}}{2a^{2}}\right) =%
\frac{1}{a^{3/2}}\frac{d^{2}a^{3/2}}{d\tau ^{2}}\equiv \frac{9}{4}f\left(
\varkappa \right)  \label{bog-1-b}
\end{equation}%
with $x=\left( \tau -\tau _{1}\right) /\tau _{0}$ where $\tau _{0}$ is the
time the transition lasts. We assume $\tau _{0}~\tau _{1}\ll 1$. In terms of 
$x$ eq. (\ref{bog-1-b}) gives%
\begin{equation}
\frac{d^{2}a^{3/2}}{dx^{2}}=\frac{9}{4}\tau _{0}^{2}a^{3/2}f\left( x\right)
\label{bog-1-e}
\end{equation}%
which can be integrated giving%
\begin{equation}
a^{3/2}\left( x\right) =A+Bx+\frac{9}{4}\tau
_{0}^{2}\int_{-x_{1}}^{x}dy\left( x-y\right) f\left( y\right) a^{3/2}\left(
y\right)  \label{bog-1-f}
\end{equation}%
where $x_{1}=\tau _{1}/\tau _{0}$. The constants of integration are obtained
by matching to the inflationary solution at $\tau =0$. We get $A=1+\tau
_{0}x_{1}$ and $B=\tau _{0}$, then%
\begin{equation}
a^{3/2}\left( x\right) =1+\tau _{0}\left( x_{1}+x\right) +\frac{9}{4}\tau
_{0}^{2}\int_{-x_{1}}^{x}dy\left( x-y\right) f\left( y\right) a^{3/2}\left(
y\right) \equiv F\left( x\right) +\tau _{0}xG\left( x\right)  \label{bog-1-g}
\end{equation}%
with%
\begin{align}
F\left( x\right) & =1+\tau _{0}x_{1}-\frac{9}{4}\tau
_{0}^{2}\int_{-x_{1}}^{x}dyyf\left( y\right) a^{3/2}\left( y\right)
\label{bog-1-h} \\
G\left( x\right) & =1+\frac{9}{4}\tau _{0}\int_{-x_{1}}^{x}dyf\left(
y\right) a^{3/2}\left( y\right)  \label{bog-1-i}
\end{align}%
For $x\geq 1$ we have $f\sim 0$, hence $F$ and $G$ are constants in that $x$
range. We now write K.G. equation (\ref{bog-1-a}) as%
\begin{equation}
\frac{d^{2}\psi _{\kappa }}{dx^{2}}+\tau _{0}^{2}\Omega _{\kappa }^{2}\left(
x\right) \psi _{\kappa }=0  \label{bog-1-j}
\end{equation}%
with%
\begin{equation}
\Omega _{\kappa }^{2}\left( x\right) =\frac{\kappa ^{2}}{a^{2}}+\frac{m^{2}}{%
H^{2}}-\frac{9}{4}f\left( x\right)  \label{bog-1-k}
\end{equation}%
We are interested in the behavior of the solutions equivalent to $H_{\nu
}^{\left( 1\right) }\left( x\right) $ (the positive frequency solutions for
a spatially flat de Sitter spacetime) in $x=-x_{1}$. There are two possible
situations. Given that $x=0$ is the middle of the transition, we consider:
(a) $\Omega \left( 0\right) \tau _{0}<1$: the details of the transition are
not important; (b) $\Omega \left( 0\right) \tau _{0}>1$: the details of the
transition matter. For modes inside the horizon ($\kappa \geq 1$) we can
consider the WKB solution%
\begin{equation}
\varphi _{\kappa +}\left( x\right) =\frac{e^{i\tau _{0}S\left[ x\right] }}{%
\sqrt{2\Omega _{\kappa }\left( x\right) }}  \label{bog-1-l}
\end{equation}%
with $\Omega \left( x\right) =dS\left[ x\right] /dx$. The derivatives are%
\begin{equation}
\frac{d\varphi _{\kappa +}}{dx}=-\left[ i\tau _{0}\Omega _{\kappa }\left(
x\right) +\frac{1}{2}\frac{\Omega _{\kappa }^{\prime }\left( x\right) }{%
\Omega _{\kappa }\left( x\right) }\right] \varphi _{\kappa +}
\label{bog-1-m}
\end{equation}%
\begin{equation}
\frac{d^{2}\varphi _{\kappa +}}{dx^{2}}=-\left[ \tau _{0}^{2}\Omega _{\kappa
}^{2}\left( x\right) -\frac{1}{4}\left( \frac{\Omega _{\kappa }^{\prime
}\left( x\right) }{\Omega _{\kappa }\left( x\right) }\right) ^{2}+\frac{1}{2}%
\left( \frac{\Omega _{\kappa }^{\prime }\left( x\right) }{\Omega _{\kappa
}\left( x\right) }\right) ^{\prime }\right] \varphi _{\kappa +}
\label{bog-1-n}
\end{equation}%
and then the equation for $\psi _{\kappa }$ reads%
\begin{align}
\frac{d^{2}\psi _{\kappa }}{dx^{2}}+&\left[ \tau _{0}^{2}\Omega _{\kappa
}^{2}\left( x\right) -\frac{1}{4}\left( \frac{\Omega _{\kappa }^{\prime
}\left( x\right) }{\Omega _{\kappa }\left( x\right) }\right) ^{2}+\frac{1}{2}%
\left( \frac{\Omega _{\kappa }^{\prime }\left( x\right) }{\Omega _{\kappa
}\left( x\right) }\right) ^{\prime }\right] \psi _{\kappa }\nonumber\\
& =\left[ -\frac{1}{%
4}\left( \frac{\Omega _{\kappa }^{\prime }\left( x\right) }{\Omega _{\kappa
}\left( x\right) }\right) ^{2}+\frac{1}{2}\left( \frac{\Omega _{\kappa
}^{\prime }\left( x\right) }{\Omega _{\kappa }\left( x\right) }\right)
^{\prime }\right] \psi _{\kappa }  \label{bog-1-o}
\end{align}%
The solution can be expressed as a superposition of positive and negative
frequency modes as%
\begin{align}
\psi _{\kappa }\left( x\right) & =\varphi _{\kappa +}\left( x\right) +\frac{i%
}{\tau _{0}}\varphi _{\kappa +}\left( x\right)
\int_{-x_{1}}^{x_{1}}dy\varphi _{\kappa -}\left( y\right) \left[ -\frac{1}{4}%
\left( \frac{\Omega _{\kappa }^{\prime }}{\Omega _{\kappa }}\right) ^{2}+%
\frac{1}{2}\left( \frac{\Omega _{\kappa }^{\prime }}{\Omega _{\kappa }}%
\right) ^{\prime }\right] \psi _{\kappa }\left( y\right)  \notag \\
& -\frac{i}{\tau _{0}}\varphi _{\kappa -}\left( x\right)
\int_{-x_{1}}^{x_{1}}dy\varphi _{\kappa +}\left( y\right) \left[ -\frac{1}{4}%
\left( \frac{\Omega _{\kappa }^{\prime }}{\Omega _{\kappa }}\right) ^{2}+%
\frac{1}{2}\left( \frac{\Omega _{\kappa }^{\prime }}{\Omega _{\kappa }}%
\right) ^{\prime }\right] \psi _{\kappa }\left( y\right)  \label{bog-1-p}
\end{align}%
Whereby we read the Bogoliubov coefficients%
\begin{equation}
\alpha _{\kappa }=1+\frac{i}{\tau _{0}}\int_{-x_{1}}^{x_{1}}dy\varphi
_{\kappa -}\left( y\right) \left[ -\frac{1}{4}\left( \frac{\Omega _{\kappa
}^{\prime }}{\Omega _{\kappa }}\right) ^{2}+\frac{1}{2}\left( \frac{\Omega
_{\kappa }^{\prime }}{\Omega _{\kappa }}\right) ^{\prime }\right] \psi
_{\kappa }\left( y\right)  \label{bog-alfa-1}
\end{equation}%
\begin{equation}
\beta _{\kappa }=-\frac{i}{\tau _{0}}\int_{-x_{1}}^{x_{1}}dy\varphi _{\kappa
+}\left( y\right) \left[ -\frac{1}{4}\left( \frac{\Omega _{\kappa }^{\prime }%
}{\Omega _{\kappa }}\right) ^{2}+\frac{1}{2}\left( \frac{\Omega _{\kappa
}^{\prime }}{\Omega _{\kappa }}\right) ^{\prime }\right] \psi _{\kappa
}\left( y\right)  \label{bog-beta-1}
\end{equation}%
which should satisfy $\left\vert \alpha _{\kappa }\right\vert
^{2}-\left\vert \beta _{\kappa }\right\vert ^{2}=1$. To obtain a simpler
expression, we consider an iterative solution. To lowest order, i.e., $\psi
\left( y\right) \simeq \varphi _{+}\left( y\right) $, we have%
\begin{equation}
\alpha _{\kappa }^{\left( 0\right) }=1+\frac{i}{\tau _{0}}%
\int_{-x_{1}}^{x_{1}}dy\varphi _{\kappa -}\left( y\right) \varphi _{\kappa
+}\left( y\right) \left[ -\frac{1}{4}\left( \frac{\Omega _{\kappa }^{\prime }%
}{\Omega _{\kappa }}\right) ^{2}+\frac{1}{2}\left( \frac{\Omega _{\kappa
}^{\prime }}{\Omega _{\kappa }}\right) ^{\prime }\right]  \label{bog-alfa-2}
\end{equation}%
\begin{equation}
\beta _{\kappa }^{\left( 0\right) }\simeq -\frac{i}{\tau _{0}}%
\int_{-x_{1}}^{x_{1}}dy\varphi _{\kappa +}^{2}\left( y\right) \left[ -\frac{1%
}{4}\left( \frac{\Omega _{\kappa }^{\prime }}{\Omega _{\kappa }}\right) ^{2}+%
\frac{1}{2}\left( \frac{\Omega _{\kappa }^{\prime }}{\Omega _{\kappa }}%
\right) ^{\prime }\right]  \label{bog-1-r}
\end{equation}%
and%
\begin{equation}
\psi _{\kappa }\left( x\right) =\alpha _{\kappa }^{\left( 0\right) }\varphi
_{\kappa +}\left( x\right) +\beta _{\kappa }^{\left( 0\right) }\varphi
_{\kappa -}\left( x\right)  \label{bog-alfa-3}
\end{equation}%
Performing another iteration we obtain%
\begin{equation}
\alpha _{\kappa }^{\left( 1\right) }\simeq \exp \left[ \frac{i}{\tau _{0}}%
\int_{-x_{1}}^{x_{1}}dy\varphi _{\kappa +}\left( y\right) \varphi _{\kappa
-}\left( y\right) Q_{\kappa }\left( y\right) \right] \left( 1+\frac{1}{2}%
\left\vert \beta _{\kappa }^{\left( 0\right) }\right\vert ^{2}\right)
\label{bog-alfa-9}
\end{equation}%
Observe that it is not necessary to perform another iteration for $\beta
_{k} $. The normalization condition $\left\vert \alpha _{\kappa }\right\vert
^{2}-\left\vert \beta _{\kappa }\right\vert ^{2}=1$ is satisfied up to a
term $\left\vert \beta _{\kappa }^{\left( 0\right) }\right\vert ^{4}/4$,
indicating that this coefficient must be $\left\vert \beta _{\kappa
}^{\left( 0\right) }\right\vert ^{2}\ll 1$ in order to render our
expressions valid.

Integrating by parts in eq eq. (\ref{bog-beta-1}) and neglecting surface
terms we have%
\begin{align}
\int_{-x_{1}}^{x_{1}}dy\varphi _{\kappa +}^{2}\left( y\right) \frac{1}{2}%
\left( \frac{\Omega _{\kappa }^{\prime }}{\Omega _{\kappa }}\right) ^{\prime
}& =-\int_{-x_{1}}^{x_{1}}dy\frac{\Omega _{\kappa }^{\prime }}{\Omega
_{\kappa }}\varphi _{\kappa +}\left( y\right) \varphi _{\kappa +}^{\prime
}\left( y\right)  \label{bog-1-s} \\
& =-\int_{-x_{1}}^{x_{1}}dy\frac{\Omega _{\kappa }^{\prime }}{\Omega
_{\kappa }}\varphi _{\kappa +}^{2}\left( y\right) \left[ -i\tau _{0}\Omega
_{\kappa }-\frac{1}{2}\frac{\Omega _{\kappa }^{\prime }}{\Omega _{\kappa }}%
\right]  \notag
\end{align}%
Observe that the first term is suppressed by a factor of $\tau _{0}$, so we
shall not consider it further. We are now at the point where details begin
to matter. Write%
\begin{equation}
\frac{\Omega _{\kappa }^{\prime }}{\Omega _{\kappa }}=\frac{1}{2\Omega
_{\kappa }^{2}}\left[ \frac{-2\kappa ^{2}}{a^{2}}\frac{a^{\prime }}{a}-\frac{%
9}{4}f^{\prime }\left( x\right) \right]  \label{bog-1-t}
\end{equation}%
The $\kappa ^{2}$ term may be neglected even when $\kappa ^{2}$ is large. To
see this, observe that%
\begin{equation}
\frac{\kappa ^{2}}{a^{2}}=\Omega _{\kappa }^{2}-\left( \frac{m}{H}\right)
^{2}+\frac{9}{4}f\left( x\right)  \label{bog-1-u}
\end{equation}%
so%
\begin{equation}
\frac{\Omega _{\kappa }^{\prime }}{\Omega _{\kappa }}=-\left( \frac{%
a^{\prime }}{a}\right) -\frac{1}{2\Omega _{\kappa }^{2}}\left[ 2\left( \frac{%
9}{4}f\left( x\right) -\left( \frac{m}{H}\right) ^{2}\right) \frac{a^{\prime
}}{a}+\frac{9}{4}f^{\prime }\left( x\right) \right]  \label{bog-1-v}
\end{equation}%
Writing%
\begin{equation}
\frac{a^{\prime }}{a}=\frac{2}{3}\frac{\left( a^{3/2}\right) ^{\prime }}{%
a^{3/2}}=\frac{2}{3}\frac{\tau _{0}G\left( x\right) }{\left( F\left(
x\right) +\tau _{0}xG\left( x\right) \right) }  \label{bog-1-w}
\end{equation}%
we see that this function is suppressed by $\tau _{0}$. Therefore in eq. (%
\ref{bog-1-v}) the only term that is not suppressed is the last one. So we
finally have%
\begin{equation}
\beta _{\kappa }^{\left( 0\right) }\sim -\frac{i}{\tau _{0}}\left( \frac{9}{%
16}\right) ^{2}\int_{-x_{1}}^{x_{1}}dy\left[ \frac{f^{\prime }\left(
y\right) }{\Omega _{\kappa }^{2}\left( y\right) }\right] ^{2}\varphi
_{\kappa +}^{2}\left( y\right)  \label{bog-1-x}
\end{equation}

We see that $\beta _{\kappa }$ is essentially the Fourier transform of $%
\left( f^{\prime }\right) ^{2}$. Since this function has a peak whose width
is $\sim 1$, by Heisenberg's principle we expect to get a negligible result
for $\tau _{0}\Omega _{\kappa }\gg 1$, namely for $\kappa \gg \tau _{0}^{-1}$%
. Observe however that this scale can be extremely high. To give concrete
results, let us consider $f^{\prime }=const=-1/2$ and assume that we can
make a linear approximation in the exponent of eq. (\ref{bog-1-l}), $S\left[
x\right] \sim S\left[ 0\right] +\Omega \left[ 0\right] x$. Assuming $%
x_{1}\simeq 1$ and that $\Omega _{\kappa }\left( y\right) $ is a slowly
varying function of time to keep only the surface terms in the integral, we
obtain eq. (\ref{bog-sw-0}). We see that the number of created particles
with large $\kappa $ is very sensitive to the details of the transition
between inflation and reheating; it would actually diverge in the limit $%
\tau_0\mapsto 0$, which is therefore unphysical.

For small $\kappa $ an instantaneous transition can be considered, and the
coefficients calculated by directly matching the inflationary and reheating
solutions at $\tau =0$. Assuming again a WKB form for the modes during
reheating and the usual Hankel function for de Sitter \cite{birrel-davies}.
For this transition the full expression for $\beta _{\kappa }$ is 
\begin{equation}
\beta _{\kappa }=-\frac{\pi ^{1/2}}{8^{1/2}\Omega _{\kappa }^{1/2}\left(
0\right) }\left\{ \kappa H_{\nu -1}^{\left( 1\right) }\left( \kappa \right)
+H_{\nu }^{\left( 1\right) }\left( \kappa \right) \left[ \frac{2}{3}\frac{%
\kappa ^{2}}{\Omega _{\kappa }^{2}\left( 0\right) }-\nu -i\Omega _{\kappa
}\left( 0\right) \right] \right\}  \label{bog-1-ad}
\end{equation}%
with $\nu =\sqrt{9/4-m^{2}/H^{2}}\simeq 3/2-m^{2}/3H^{2}$. For $\kappa <1$
the Hankel functions can be approximated as%
\begin{equation}
H_{\nu }^{\left( 1\right) }\left( \kappa \right) \simeq -\frac{i}{\pi }%
\Gamma \left( \nu \right) \left( \frac{\kappa }{2}\right) ^{-\nu }
\label{bog-1-ae}
\end{equation}%
Using $\left( \nu -1\right) \Gamma \left( \nu -1\right) =\Gamma \left( \nu
\right) $ and the fact that $2\kappa ^{2}/\Omega _{\kappa }^{2}\left(
0\right) 3-\nu \sim 1$ we get eq. (\ref{bog-lw})

\section{Calculation of $\left\langle \protect\rho+p\right\rangle $ and $n$}

In terms of the scalar field we have%
\begin{equation}
\left\langle \rho+p\right\rangle \approx\frac{4\left\langle
T_{\Phi}^{00}\right\rangle +\left\langle T_{\Phi a}^{a}\right\rangle }{3}
\label{rho-1}
\end{equation}
Using eq. (\ref{c}) we obtain%
\begin{equation}
4\left\langle T_{\Phi}^{00}\right\rangle +\left\langle T_{\Phi
a}^{a}\right\rangle =H^{4}\left[ 3\left\langle \dot{\Phi}\dot{\Phi}^{\dagger
}\right\rangle +\frac{1}{a^{2}}\left\langle \bar{\nabla}\Phi\cdot\bar{\nabla 
}\Phi^{\dagger}\right\rangle \right]  \label{rho-6}
\end{equation}
Replacing the decompositions (\ref{jb}) and using 
\begin{equation}
\left[ a_{\kappa},a_{\varpi}^{\dagger}\right] =\left( 2\pi\right)
^{3/2}a^{3}\left( \tau\right) \delta\left( \bar{\kappa}-\bar{\varpi }\right)
\label{rho-11}
\end{equation}
we obtain%
\begin{eqnarray}
3\left\langle \dot{\Phi}\dot{\Phi}^{\dagger}\right\rangle +\frac{1}{a^{2}}%
\left\langle \bar{\nabla}\Phi\cdot\bar{\nabla}\Phi^{\dagger}\right\rangle &=&%
\frac{1}{\left( 2\pi\right) ^{3/2}a^{3}\left( \tau\right) }\int d\bar{\kappa}%
\left\{ 3\dot{\phi}_{\kappa}^{I}\left( \tau\right) \dot{\phi }%
_{\kappa}^{I\ast}\left( \tau\right) \right.  \notag \\
&-& 9\frac{\dot{a}\left( \tau\right) }{a\left( \tau\right) }\left[ \dot{\phi}%
_{\kappa}^{I}\left( \tau\right) \phi_{\kappa}^{I\ast}\left( \tau\right)
+\phi_{\kappa}^{I}\left( \tau\right) \dot{\phi}_{\kappa}^{I\ast}\left(
\tau\right) \right]  \label{rho-14} \\
&+& \left.\left[27\frac{\dot{a}^{2}\left( \tau\right) }{a^{2}\left(
\tau\right) }+\frac{\kappa^{2}}{a^{2}\left( \tau\right) }\right]
\phi_{\kappa}^{I}\left( \tau\right) \phi_{\kappa}^{I\ast}\left( \tau\right)
\right\}  \notag
\end{eqnarray}
Using decomposition (\ref{corr-6}) we identify two different contributions
to the integrand: one from pure vacuum%
\begin{eqnarray}
M_{0}&=&3\dot{\phi}_{k}^{R}\left( \tau\right) \dot{\phi}_{k}^{R\ast}\left(
\tau\right) -9\frac{\dot{a}\left( \tau\right) }{a\left( \tau\right) }\left[ 
\dot{\phi}_{k}^{R}\left( \tau\right) \phi_{k}^{R\ast}\left( \tau\right)
+\phi_{k}^{R}\left( \tau\right) \dot{\phi}_{k}^{R\ast}\left( \tau\right) %
\right]  \notag \\
&+& 27\frac{\dot{a}^{2}\left( \tau\right) }{a^{2}\left( \tau\right) }%
\phi_{k}^{R}\left( \tau\right) \phi_{k}^{R\ast}\left( \tau\right) +\frac{%
\kappa^{2}}{a^{2}\left( \tau\right) }\phi_{k}^{R}\left( \tau\right)
\phi_{k}^{R\ast}\left( \tau\right)  \label{rho-14-a}
\end{eqnarray}
and one from the created particles,%
\begin{align}
M_{1} & =3\left[ \alpha_{k}\beta_{k}^{\ast}\dot{\phi}_{k}^{R}\left(
\tau\right) \dot{\phi}_{k}^{R}\left( \tau\right) +\alpha_{k}^{\ast}\beta_{k}%
\dot{\phi}_{k}^{R\ast}\left( \tau\right) \dot{\phi}_{k}^{R\ast }\left(
\tau\right) \right]  \notag \\
& -18\frac{\dot{a}\left( \tau\right) }{a\left( \tau\right) }\left[
\alpha_{k}\beta_{k}^{\ast}\phi_{k}^{R}\left( \tau\right) \dot{\phi}%
_{k}^{R}\left( \tau\right) +\alpha_{\kappa}^{\ast }\beta_{\kappa}\dot{\phi}%
_{k}^{R\ast}\left( \tau\right) \phi_{k}^{R\ast }\left( \tau\right) \right] 
\notag \\
& +\left( 27\frac{\dot{a}^{2}\left( \tau\right) }{a^{2}\left( \tau\right) }+%
\frac{\kappa^{2}}{a^{2}\left( \tau\right) }\right) \left[ \alpha_{k}%
\beta_{k}^{\ast}\phi_{k}^{R2}\left( \tau\right) +\alpha_{k}^{\ast
}\beta_{k}\phi_{k}^{R\ast2}\left( \tau\right) \right]  \label{rho-14-b} \\
& +2\left\vert \beta_{k}\right\vert ^{2}\left[ 3\dot{\phi}_{k}^{R\ast
}\left( \tau\right) \dot{\phi}_{k}^{R}\left( \tau\right) -9\frac{\dot {a}%
\left( \tau\right) }{a\left( \tau\right) }\left( \dot{\phi}_{k}^{R}\left(
\tau\right) \phi_{k}^{R\ast}\left( \tau\right) +\dot{\phi}_{k}^{R\ast}\left(
\tau\right) \phi_{k}^{R}\left( \tau\right) \right)\right.  \notag \\
& +\left. \left( 27\frac{\dot{a}^{2}\left( \tau\right) }{a^{2}\left(
\tau\right) }+\frac{\kappa^{2}}{a^{2}\left( \tau\right) }\right)
\phi_{k}^{R\ast }\left( \tau\right) \phi_{k}^{R}\left( \tau\right) \right] 
\notag
\end{align}
We are interested in $M_{1}$ and of it, the contribution from the $%
\left\vert \beta_{k}\right\vert ^{2}$ terms is the most important: as $%
\alpha_{k}$ oscillates (see eq, [\ref{bog-alfa-9}]), the terms proportional
to $\alpha _{k}$ and $\alpha_{k}^{\ast}$ will give negligible contributions
when integrated. Replacing the WKB form for $\phi_{k}^{R}\left( \tau\right)$%
, eq. (\ref{k-h-b}) we obtain

\begin{align}
3\left\langle \dot{\Phi}\dot{\Phi}^{\dagger }\right\rangle +\frac{1}{a^{2}}%
\left\langle \bar{\nabla}\Phi \cdot \bar{\nabla}\Phi ^{\dagger
}\right\rangle & =\frac{1}{\left( 2\pi \right) ^{3/2}a^{3}\left( \tau
\right) }\int d\bar{\kappa}\left\vert \beta _{k}\right\vert ^{2}\left\{ 3%
\left[ \Omega _{k}\left( \tau \right) +\frac{1}{4}\frac{\dot{\Omega}%
_{k}^{2}\left( \tau \right) }{\Omega _{k}^{3}\left( \tau \right) }\right]
\right.  \notag \\
& +\left. \frac{9}{2}\frac{\dot{a}\left( \tau \right) }{a\left( \tau \right) 
}\frac{\dot{\Omega}_{k}\left( \tau \right) }{\Omega _{k}^{2}\left( \tau
\right) }+\left[ 27\frac{\dot{a}^{2}\left( \tau \right) }{a^{2}\left( \tau
\right) }+\frac{\kappa ^{2}}{a^{2}\left( \tau \right) }\right] \frac{1}{%
\Omega _{k}\left( \tau \right) }\right\}  \label{rho-25}
\end{align}%
Replacing%
\begin{align}
\frac{\dot{a}}{a} &=\frac{1}{a^{3/2}},\nonumber\\
\dot{\Omega}_{\kappa }\left( \tau\right) &=-\frac{1}{a^{7/2}}
\frac{\kappa ^{2}}{\Omega _{\kappa }\left( \tau
\right) },\nonumber\\
\frac{\dot{\Omega}_{k}\left( \tau \right) }{\Omega
_{k}^{2}\left( \tau \right) }&=-\frac{1}{a^{7/2}}\frac{\kappa ^{2}}{\Omega
_{\kappa }^{3}\left( \tau \right) },\nonumber\\
\frac{\dot{\Omega}_{k}^{2}\left(
\tau \right) }{\Omega _{k}^{3}\left( \tau \right) }& =\frac{1}{a^{7}}\frac{%
\kappa ^{4}}{\Omega _{\kappa }^{5}\left( \tau \right) }  \label{rho-26}
\end{align}%
we have that 
\begin{align}
3\left\langle \dot{\Phi}\dot{\Phi}^{\dagger }\right\rangle +&\frac{1}{a^{2}}%
\left\langle \bar{\nabla}\Phi \cdot \bar{\nabla}\Phi ^{\dagger
}\right\rangle  \rightarrow \nonumber\\
& \rightarrow \frac{1}{\left( 2\pi \right) ^{3/2}a^{3}\left(
 \tau \right) }\int d\bar{\kappa}\left\vert \beta _{k}\right\vert ^{2}
\left\{3\left[ \Omega _{k}\left( \tau \right) +\frac{1}{4}\frac{\kappa ^{4}}{%
a^{7}\left( \tau \right) \Omega _{\kappa }^{5}\left( \tau \right) }\right]
\right.  \notag \\
& -\left. \frac{9}{2}\frac{\kappa ^{2}}{a^{5}\left( \tau \right) \Omega
_{\kappa }^{3}\left( \tau \right) }+\left[ \frac{27}{a^{3}\left( \tau
\right) }+\frac{\kappa ^{2}}{a^{2}\left( \tau \right) }\right] \frac{1}{%
\Omega _{k}\left( \tau \right) }\right\}  \label{rho-27}
\end{align}%
By simple inspection we can see that the terms that contribute the most are%
\begin{align}
3\left\langle \dot{\Phi}\dot{\Phi}^{\dagger }\right\rangle & +\frac{1}{a^{2}}%
\left\langle \bar{\nabla}\Phi \cdot \bar{\nabla}\Phi ^{\dagger
}\right\rangle \simeq \nonumber\\
&\simeq \frac{1}{\left( 2\pi \right) ^{3/2}a^{3}\left( \tau
\right) }\int d\bar{\kappa}\left\vert \beta _{k}\right\vert ^{2}\left[
3\Omega _{k}\left( \tau \right) +\frac{\kappa ^{2}}{a^{2}\left( \tau \right) 
}\frac{1}{\Omega _{k}\left( \tau \right) }\right]  \label{rho-28}
\end{align}
because they decay more slowly than the others. We must now replace the
Bogoliubov coefficients and perform the integrations. For long wavelengths,
i.e., those in the the interval $\left( 0,1\right) $ we use the expression (%
\ref{bog-lw}). Thus in this case we must evaluate%
\begin{align}
 3\left\langle \dot{\Phi}\dot{\Phi}^{\dagger }\right\rangle &+\left.\frac{1}{%
a^{2}}\left\langle \bar{\nabla}\Phi \cdot \bar{\nabla}\Phi ^{\dagger
}\right\rangle \right\vert _{\left( 0\right) }\nonumber\\
& \simeq \frac{\left( \nu -1\right) ^{2}\Gamma ^{2}\left( 1/2\right) }
{\left( 2\pi \right) ^{3/2}\pi
a^{3}\left( \tau \right) }\int_{0}^{1}\frac{d\bar{\kappa}}{\Omega _{\kappa
}\left( 0\right) }\frac{1}{\kappa ^{2\nu }}\left[ 3\Omega _{k}\left( \tau
\right) +\frac{\kappa ^{2}}{a^{2}\left( \tau \right) }\frac{1}{\Omega
_{k}\left( \tau \right) }\right]  \label{rho-28-b}
\end{align}%
Since $m/H\ll 1$ and we are considering a period of time in which $a\left(
\tau \right) $ does not differ very much from unity, we can take $a\left(
\tau \right) \simeq 1$ in all the roots, and so we have%
\begin{align}
\left. 3\left\langle \dot{\Phi}\dot{\Phi}^{\dagger }\right\rangle +\frac{1}{%
a^{2}}\left\langle \bar{\nabla}\Phi \cdot \bar{\nabla}\Phi ^{\dagger
}\right\rangle \right\vert _{\left( 0\right) }& \simeq \frac{1}{2\left( 2\pi
\right) ^{1/2}a^{4}\left( \tau \right) }\nonumber\\
& \times \int_{0}^{1}d\kappa \left[ \frac{3}{%
\kappa ^{2\nu -2}}+\frac{\kappa ^{4-2\nu }}{\left( \kappa
^{2}+m^{2}/H^{2}\right) }\right]  \label{rho-29}
\end{align}%
And finally the contribution from long wavelengths reads 
\begin{equation}
\left. 3\left\langle \dot{\Phi}\dot{\Phi}^{\dagger }\right\rangle +\frac{1}{%
a^{2}}\left\langle \bar{\nabla}\Phi \cdot \bar{\nabla}\Phi ^{\dagger
}\right\rangle \right\vert _{\left( 0\right) }\simeq \frac{1}{\left( 2\pi
\right) ^{1/2}a^{4}\left( \tau \right) }\frac{3}{4}\left( \frac{H}{m}\right)
^{2}  \label{rho-30}
\end{equation}%
To evaluate the contribution from the short wavelengths we use eq. (\ref%
{bog-sw}) in eq. (\ref{rho-28}), so we have to compute

\begin{align}
3\left\langle \dot{\Phi}\dot{\Phi}^{\dagger }\right\rangle + &\left.\frac{1}{%
a^{2}}\left\langle \bar{\nabla}\Phi \cdot \bar{\nabla}\Phi ^{\dagger
}\right\rangle \right\vert _{\left( \infty \right) } \simeq \left( \frac{9}{%
16}\right) ^{4}\frac{1}{16}\frac{1}{\tau _{0}^{4}}\frac{2}{\left( 2\pi
\right) ^{1/2}a^{4}\left( \tau \right) }  \label{rho-31} \\
& \times \int_{1}^{\infty }d\kappa \frac{1}{\kappa ^{8}}\left[ 3\left(
\kappa ^{2}+a^{2}\left( \tau \right) m^{2}/H^{2}\right) ^{1/2}+\frac{\kappa
^{2}}{\left( \kappa ^{2}+a^{2}\left( \tau \right) m^{2}/H^{2}\right) ^{1/2}}%
\right]  \notag
\end{align}%
As in this case $m/H\ll \kappa $, we can neglect that term and so the
contribution from short wavelengths reads%
\begin{equation}
\left. 3\left\langle \dot{\Phi}\dot{\Phi}^{\dagger }\right\rangle +\frac{1}{%
a^{2}}\left\langle \bar{\nabla}\Phi \cdot \bar{\nabla}\Phi ^{\dagger
}\right\rangle \right\vert _{\left( \infty \right) }\simeq \left( \frac{9}{16%
}\right) ^{4}\frac{1}{24}\frac{1}{\tau _{0}^{2}}\frac{\pi }{\left( 2\pi
\right) ^{3/2}a^{4}\left( \tau \right) }  \label{rho-32}
\end{equation}%
Gathering expressions (\ref{rho-30}) and (\ref{rho-32}), approximating $\nu
-1\simeq 1/2$ and using that $\Gamma \left( 1/2\right) =\pi ^{1/2}$ we
arrive at expression (\ref{rho-p}).

For $n$, the number density of created particles, we have \cite%
{birrel-davies,MukWin07,ParTom09}%
\begin{equation}
n=\frac{H^{3}}{a^{3}\left( \tau \right) }\int_{0}^{\infty }d\bar{\kappa}%
\left\vert \beta _{\kappa }\right\vert ^{2}  \label{n-1}
\end{equation}%
Using again eqs. (\ref{bog-lw}) and (\ref{bog-sw}) in the appropriate
momentum intervals we obtain eq. (\ref{2c}).

\section{Calculation of the velocity correlation spectrum}

We start by replacing eq. (\ref{corr-6-a}) into (\ref{corr-5}) and noting
that three kernels build the correlation function: one with the vacuum
contributions%
\begin{align}
N_{\left( 00\right) }^{ij}\left( \kappa ,\varpi ,\tau ,\tau ^{\prime
}\right) & =\varpi ^{i}\kappa ^{j}\frac{\partial }{\partial \tau }G_{\kappa
}^{R+}\left( \tau ,\tau ^{\prime }\right) \frac{\partial }{\partial \tau
^{\prime }}G_{\varpi }^{R+}\left( \tau ,\tau ^{\prime }\right) \nonumber\\
& +\kappa^{i}\varpi ^{j}\frac{\partial }{\partial \tau }G_{\varpi }^{R+}\left( \tau
,\tau ^{\prime }\right) \frac{\partial }{\partial \tau ^{\prime }}G_{\kappa
}^{R+}\left( \tau ,\tau ^{\prime }\right)  \label{kerr-00} \\
& +\varpi ^{i}\varpi ^{j}\left[ \frac{\partial ^{2}}{\partial \tau ^{\prime
}\partial \tau }G_{\kappa }^{R+}\left( \tau ,\tau ^{\prime }\right) \right]
G_{\varpi }^{R+}\left( \tau ,\tau ^{\prime }\right) \nonumber\\
& +\kappa ^{i}\kappa ^{j}%
\left[ \frac{\partial ^{2}}{\partial \tau ^{\prime }\partial \tau }G_{\varpi
}^{R+}\left( \tau ,\tau ^{\prime }\right) \right] G_{\kappa }^{R+}\left(
\tau ,\tau ^{\prime }\right)  \nonumber
\end{align}%
another with mixed contributions from the vacuum and the created particles,%
\begin{align}
&N_{\left( 01\right) }^{ij}\left( \kappa ,\varpi ,\tau ,\tau ^{\prime
}\right) =\varpi ^{i}\kappa ^{j}\left\vert \beta _{\kappa }\right\vert
^{2} \left[ \frac{\partial }{\partial \tau }G_{\kappa }^{R+}\left( \tau
,\tau ^{\prime }\right) +\frac{\partial }{\partial \tau }G_{\kappa
}^{R-}\left( \tau ,\tau ^{\prime }\right) \right] \frac{\partial }{\partial
\tau ^{\prime }}G_{\varpi }^{R+}\left( \tau ,\tau ^{\prime }\right)
\nonumber \\
& +\varpi ^{i}\kappa ^{j}\left\vert \beta _{\varpi }\right\vert ^{2}\frac{%
\partial }{\partial \tau }G_{\kappa }^{R+}\left( \tau ,\tau ^{\prime
}\right) \left[ \frac{\partial }{\partial \tau ^{\prime }}G_{\varpi
}^{R+}\left( \tau ,\tau ^{\prime }\right) +\frac{\partial }{\partial \tau
^{\prime }}G_{\varpi }^{R-}\left( \tau ,\tau ^{\prime }\right) \right] 
\label{ker-01} \\
& +\varpi ^{i}\varpi ^{j}\left\vert \beta _{\kappa }\right\vert ^{2}\left[ 
\frac{\partial ^{2}}{\partial \tau ^{\prime }\partial \tau }G_{\kappa
}^{R+}\left( \tau ,\tau ^{\prime }\right) +\frac{\partial ^{2}}{\partial
\tau ^{\prime }\partial \tau }G_{\kappa }^{R-}\left( \tau ,\tau ^{\prime
}\right) \right] G_{\varpi }^{R+}\left( \tau ,\tau ^{\prime }\right)  \notag
\\
& +\varpi ^{i}\varpi ^{j}\left\vert \beta _{\varpi }\right\vert ^{2}\left[ 
\frac{\partial ^{2}}{\partial \tau ^{\prime }\partial \tau }G_{\kappa
}^{R+}\left( \tau ,\tau ^{\prime }\right) \right] \left[ G_{\varpi
}^{R+}\left( \tau ,\tau ^{\prime }\right) +G_{\varpi }^{R-}\left( \tau ,\tau
^{\prime }\right) \right]  \notag \\
& +\varpi ^{i}\kappa ^{j}\alpha _{\varpi }\beta _{\varpi }^{\ast }\frac{%
\partial }{\partial \tau }G_{\kappa }^{R+}\left( \tau ,\tau ^{\prime
}\right) g_{\varpi }\left( \tau \right) \frac{\partial }{\partial \tau
^{\prime }}g_{\varpi }\left( \tau ^{\prime }\right) 
+\varpi ^{i}\kappa
^{j}\alpha _{\varpi }^{\ast }\beta _{\varpi }\frac{\partial }{\partial \tau }%
G_{\kappa }^{R+}\left( \tau ,\tau ^{\prime }\right) g_{\varpi }^{\ast
}\left( \tau \right) \frac{\partial }{\partial \tau ^{\prime }}g_{\varpi
}^{\ast }\left( \tau ^{\prime }\right)  \notag \\
& +\varpi ^{i}\kappa ^{j}\alpha _{\kappa }\beta _{\kappa }^{\ast }\frac{%
\partial }{\partial \tau }g_{\kappa }\left( \tau \right) g_{\kappa }\left(
\tau ^{\prime }\right) \frac{\partial }{\partial \tau ^{\prime }}G_{\varpi
}^{R+}\left( \tau ,\tau ^{\prime }\right)
+\varpi ^{i}\kappa ^{j}\alpha
_{\kappa }^{\ast }\beta _{\kappa }\frac{\partial }{\partial \tau }g_{\kappa
}^{\ast }\left( \tau \right) g_{\kappa }^{\ast }\left( \tau ^{\prime
}\right) \frac{\partial }{\partial \tau ^{\prime }}G_{\varpi }^{R+}\left(
\tau ,\tau ^{\prime }\right)  \notag \\
& +\varpi ^{i}\varpi ^{j}\alpha _{\kappa }\beta _{\kappa }^{\ast }\left[ 
\frac{\partial ^{2}}{\partial \tau ^{\prime }\partial \tau }g_{\kappa
}\left( \tau \right) g_{\kappa }\left( \tau ^{\prime }\right) \right]
G_{\varpi }^{R+}\left( \tau ,\tau ^{\prime }\right)
+\varpi ^{i}\varpi
^{j}\alpha _{\varpi }^{\ast }\beta _{\varpi }\left[ \frac{\partial ^{2}}{%
\partial \tau ^{\prime }\partial \tau }G_{\kappa }^{R+}\left( \tau ,\tau
^{\prime }\right) \right] g_{\varpi }^{\ast }\left( \tau \right) g_{\varpi
}^{\ast }\left( \tau ^{\prime }\right)  \notag \\
& +\varpi ^{i}\varpi ^{j}\alpha _{\kappa }^{\ast }\beta _{\kappa }\left[ 
\frac{\partial ^{2}}{\partial \tau ^{\prime }\partial \tau }g_{\kappa
}^{\ast }\left( \tau \right) g_{\kappa }^{\ast }\left( \tau ^{\prime
}\right) \right] G_{\varpi }^{R+}\left( \tau ,\tau ^{\prime }\right)
+\varpi
^{i}\varpi ^{j}\alpha _{\varpi }\beta _{\varpi }^{\ast }\left[ \frac{%
\partial ^{2}}{\partial \tau ^{\prime }\partial \tau }G_{\kappa }^{R+}\left(
\tau ,\tau ^{\prime }\right) \right] g_{\varpi }\left( \tau \right)
g_{\varpi }\left( \tau ^{\prime }\right)  \notag \\
& +\left( \varpi \leftrightarrow \kappa \right) ,  \notag
\end{align}%
and a third kernel with contributions from the created particles,%
\begin{align}
N_{\left( 11\right) }^{ij}\left( \kappa ,\varpi ,\tau ,\tau ^{\prime
}\right) & =\varpi ^{i}\kappa ^{j}\left\vert \beta _{\kappa }\right\vert
^{2}\left\vert \beta _{\varpi }\right\vert ^{2}\left( \frac{\partial }{%
\partial \tau }G_{\kappa }^{R+}\left( \tau ,\tau ^{\prime }\right) +\frac{%
\partial }{\partial \tau }G_{\kappa }^{R-}\left( \tau ,\tau ^{\prime
}\right) \right) \notag\\
& \times\left( \frac{\partial }{\partial \tau ^{\prime }}G_{\varpi
}^{R+}\left( \tau ,\tau ^{\prime }\right) +\frac{\partial }{\partial \tau
^{\prime }}G_{\varpi }^{R-}\left( \tau ,\tau ^{\prime }\right) \right) 
\notag \\
& +\varpi ^{i}\varpi ^{j}\alpha _{\kappa }\beta _{\kappa }^{\ast }\left\vert
\beta _{\varpi }\right\vert ^{2}\left( \frac{\partial ^{2}}{\partial \tau
^{\prime }\partial \tau }g_{\kappa }\left( \tau \right) g_{\kappa }\left(
\tau ^{\prime }\right) \right) \notag\\
&\times \left( G_{\varpi }^{R+}\left( \tau ,\tau
^{\prime }\right) +G_{\varpi }^{R-}\left( \tau ,\tau ^{\prime }\right)
\right)  \notag \\
& +\varpi ^{i}\kappa ^{j}\left[ \alpha _{\kappa }\beta _{\kappa }^{\ast
}\alpha _{\varpi }\beta _{\varpi }^{\ast }\frac{\partial }{\partial \tau }%
g_{\kappa }\left( \tau \right) g_{\kappa }\left( \tau ^{\prime }\right)
g_{\varpi }\left( \tau \right) \frac{\partial }{\partial \tau ^{\prime }}%
g_{\varpi }\left( \tau ^{\prime }\right) +\cdots \right]  \label{ker-11} \\
+& \varpi ^{i}\varpi ^{j}\left[ \alpha _{\kappa }^{\ast }\beta _{\kappa
}\alpha _{\varpi }\beta _{\varpi }^{\ast }\left( \frac{\partial ^{2}}{%
\partial \tau ^{\prime }\partial \tau }g_{\kappa }^{\ast }\left( \tau
\right) g_{\kappa }^{\ast }\left( \tau ^{\prime }\right) \right) g_{\varpi
}\left( \tau \right) g_{\varpi }\left( \tau ^{\prime }\right) +\cdots \right]
\notag \\
& +\left( \varpi \left. \leftrightarrow \right. \kappa \right)  \notag
\end{align}%
where the dots in square brackets indicate more terms with combinations of $%
\alpha _{\kappa }$, $\beta _{\kappa }^{\ast }$, $\alpha _{\varpi }$, $\beta
_{\varpi }^{\ast }$. Of the three kernels, $N_{\left( 11\right) }^{ij}\left(
\kappa ,\varpi ,\tau ,\tau ^{\prime }\right) $ gives the main contribution,
because it has no vacuum contribution. Observe that as the coefficient $%
\alpha _{\kappa }$ is oscillatory (see eq. [\ref{bog-alfa-9}] in Appendix
A), the terms with coefficients with $\alpha _{\kappa }$ and $\alpha
_{\varpi }$ will give negligible contributions when integrated. Therefore in
what follows we shall analyze only the terms in $\left\vert \beta _{\kappa
}\right\vert ^{2}\left\vert \beta _{\varpi }\right\vert ^{2}$. From direct
inspection of eq. (\ref{ker-11}) we see that the only terms that can survive
after integrating are those proportional to $\varpi ^{i}\varpi ^{j}$ and to $%
\kappa ^{i}\kappa ^{j}$. Thus we have to evaluate
\begin{align}
&\left\langle 0\left\vert \left\{ T_{\Phi }^{\left\{ 0i\right\} }\left(
x^{\mu }\right) ,T_{\Phi }^{\left\{ 0j\right\} }\left( x^{\prime \nu
}\right) \right\} \right\vert 0\right\rangle _{\beta } =\notag\\
& =\frac{H^{8}}{32\pi
^{3}a^{6}\left( \tau \right) }\iint d\bar{\kappa}d\bar{\varpi}~e^{i\left( 
\bar{\kappa}+\bar{\varpi}\right) .\left( \bar{r}-\bar{r}^{\prime }\right)
}\left\vert \beta _{\kappa }\right\vert ^{2}\left\vert \beta _{\varpi
}\right\vert ^{2}  \label{corr-ap-2} \\
& \times  \left\{ \varpi ^{i}\varpi ^{j}\left[ G_{\varpi }^{R+}\left( \tau
,\tau ^{\prime }\right) +G_{\varpi }^{R-}\left( \tau ,\tau ^{\prime }\right) %
\right] \left[ \frac{\partial ^{2}}{\partial \tau ^{\prime }\partial \tau }%
G_{\kappa }^{R+}\left( \tau ,\tau ^{\prime }\right) +\frac{\partial ^{2}}{%
\partial \tau ^{\prime }\partial \tau }G_{\kappa }^{R-}\left( \tau ,\tau
^{\prime }\right) \right] \right.  \notag \\
&+ \left. \kappa ^{i}\kappa ^{j}\left[ G_{\kappa }^{R+}\left( \tau ,\tau
^{\prime }\right) +G_{\kappa }^{R-}\left( \tau ,\tau ^{\prime }\right) %
\right] \left[ \frac{\partial ^{2}}{\partial \tau ^{\prime }\partial \tau }%
G_{\varpi }^{R+}\left( \tau ,\tau ^{\prime }\right) +\frac{\partial ^{2}}{%
\partial \tau ^{\prime }\partial \tau }G_{\varpi }^{R-}\left( \tau ,\tau
^{\prime }\right) \right] \right\}  \notag
\end{align}

The modes during reheating are of the WKB form, and thus the $%
G_{k}^{R+}\left( \tau ,\tau ^{\prime }\right) $ reads%
\begin{equation}
G_{k}^{R+}\left( \tau ,\tau ^{\prime }\right) =\frac{1}{2\sqrt{\Omega
_{k}\left( \tau \right) \Omega _{k}\left( \tau ^{\prime }\right) }}\exp %
\left[ -i\int_{\tau ^{\prime }}^{\tau }\Omega _{k}\left( \sigma \right)
d\sigma \right]  \label{green-pos-b}
\end{equation}%
The velocity spectrum is defined in eq. (\ref{corr-9-a}), so taking the
coincidence limit $\tau =\tau ^{\prime }$ of eq. (\ref{corr-9}) and
transforming Fourier we obtain
\begin{align}
& \Phi ^{ij}\left( \varsigma ,\tau \right)  =\frac{H^{5}}{32\pi
^{3}a^{3}\left( \tau \right) }\frac{1}{\left\langle \rho +p\right\rangle ^{2}%
}\varsigma ^{i}\varsigma ^{j} \int d\bar{\kappa}\left\vert \beta _{\kappa
}\right\vert ^{2}\left\vert \beta _{\varsigma -\kappa }\right\vert ^{2}\notag\\
&\times \left[
G_{\varsigma -\kappa }^{R+}\left( \tau ,\tau \right) +G_{\varsigma -\kappa
}^{R-}\left( \tau ,\tau \right) \right] 
 \left[ \frac{\partial ^{2}}{\partial \tau ^{\prime }\partial \tau }%
G_{\kappa }^{R+}\left( \tau ,\tau \right) +\frac{\partial ^{2}}{\partial
\tau ^{\prime }\partial \tau }G_{\kappa }^{R-}\left( \tau ,\tau \right) %
\right]  \notag \\
& +\frac{m^{4}H^{4}}{32\pi ^{3}a^{6}\left( \tau \right) }\frac{1}{%
\left\langle \rho +p\right\rangle ^{2}}\frac{\delta ^{ij}}{3}
\int d\bar{
\kappa}\left\vert \beta _{\kappa }\right\vert ^{2}\left\vert \beta
_{\varsigma -\kappa }\right\vert ^{2}\kappa ^{2} \label{corr-17}  \\
&\times \left\{ \left[ G_{\varsigma
-\kappa }^{R+}\left( \tau ,\tau \right) +G_{\varsigma -\kappa }^{R-}\left(
\tau ,\tau \right) \right] \right.  
\left[ \frac{\partial ^{2}}{\partial \tau ^{\prime }\partial \tau }%
G_{\kappa }^{R+}\left( \tau ,\tau \right) +\frac{\partial ^{2}}{\partial
\tau ^{\prime }\partial \tau }G_{\kappa }^{R-}\left( \tau ,\tau \right) %
\right]  \notag \\
& +\left. \left[ G_{\kappa }^{R+}\left( \tau ,\tau ^{\prime }\right)
+G_{\kappa }^{R-}\left( \tau ,\tau ^{\prime }\right) \right] \left[ \frac{%
\partial ^{2}}{\partial \tau ^{\prime }\partial \tau }G_{\varsigma -\kappa
}^{R+}\left( \tau ,\tau ^{\prime }\right) +\frac{\partial ^{2}}{\partial
\tau ^{\prime }\partial \tau }G_{\varsigma -\kappa }^{R-}\left( \tau ,\tau
\right) \right] \right\}  \notag
\end{align}%
where we used the isotropy of the Bogoliubov coefficients to replace 
\begin{equation}
\kappa ^{i}\kappa ^{j}\mapsto \frac{1}{3}\kappa ^{2}\delta ^{ij}
\label{corr-18}
\end{equation}%
as those terms are the ones that give non null contributions. After
replacing the propagators and their derivatives, the velocity correlation
can be written as%
\begin{equation}
\Phi ^{ij}\left( \varsigma ,\tau \right) =\Phi _{(1)}^{ij}\left( \varsigma
,\tau \right) +\Phi _{(2)}^{ij}\left( \varsigma ,\tau \right)
\label{corr-19}
\end{equation}%
with%
\begin{align}
\Phi _{(1)}^{ij}\left( \varsigma ,\tau \right) & =\frac{H^{5}}{32\pi
^{3}a\left( \tau \right) }\frac{1}{\left\langle \rho +p\right\rangle ^{2}}%
\varsigma ^{i}\varsigma ^{j}\int d\bar{\kappa}\left\vert \beta _{\kappa
}\right\vert ^{2}\left\vert \beta _{\varsigma -\kappa }\right\vert ^{2}
\label{corr-20} \\
& \times \left[ \left\vert \bar{\kappa}-\bar{\varsigma}\right\vert
^{2}+a^{2}\left( \tau \right) \left( \frac{m}{H}\right) ^{2}\right] ^{-1/2}%
\left[ \kappa ^{2}+a^{2}\left( \tau \right) \left( \frac{m}{H}\right) ^{2}%
\right] ^{-1/2}  \notag \\
\times & \left\{ \frac{1}{4}\frac{\kappa ^{4}}{a^{3}\left( \tau \right) }%
\left[ \kappa ^{2}+a^{2}\left( \tau \right) \left( \frac{m}{H}\right) ^{2}%
\right] ^{-2}+\frac{1}{a^{2}\left( \tau \right) }\left[ \kappa
^{2}+a^{2}\left( \tau \right) \left( \frac{m}{H}\right) ^{2}\right] \right\}
\notag
\end{align}%
and%
\begin{align}
\Phi _{(2)}^{ij}&\left( \zeta ,\tau \right)  =\frac{H^{5}}{32\pi ^{3}a\left(
\tau \right) }\frac{1}{\left\langle \rho +p\right\rangle ^{2}}\frac{\delta
^{ij}}{3}\int d\bar{\kappa}\kappa ^{2}\left\vert \beta _{\kappa }\right\vert
^{2}\left\vert \beta _{\kappa -\varsigma }\right\vert ^{2}  \label{corr-21}
\\
& \times \left[ \left\vert \bar{\kappa}-\bar{\varsigma}\right\vert
^{2}+a^{2}\left( \tau \right) \left( \frac{m}{H}\right) ^{2}\right] ^{-1/2}%
\left[ \kappa ^{2}+a^{2}\left( \tau \right) \left( \frac{m}{H}\right) ^{2}%
\right] ^{-1/2}  \notag \\
& \times \left\{ \frac{1}{4}\frac{\kappa ^{4}}{a^{3}\left( \tau \right) }%
\left[ \kappa ^{2}+a^{2}\left( \tau \right) \left( \frac{m}{H}\right) ^{2}%
\right] ^{-2}+\frac{1}{a^{2}\left( \tau \right) }\left[ \kappa
^{2}+a^{2}\left( \tau \right) \left( \frac{m}{H}\right) ^{2}\right] \right. 
\notag \\
& +\left. \frac{1}{4}\frac{\left\vert \bar{\kappa}-\bar{\varsigma}%
\right\vert ^{4}}{a^{3}\left( \tau \right) }\left[ \left\vert \bar{\kappa}-%
\bar{\varsigma}\right\vert ^{2}+a^{2}\left( \tau \right) \left( \frac{m}{H}%
\right) ^{2}\right] ^{-2} + \frac{1}{a^{2}\left( \tau \right) }\left[
\left\vert \bar{\kappa}-\bar{\varsigma}\right\vert ^{2}+a^{2}\left( \tau
\right) \left( \frac{m}{H}\right) ^{2}\right] \right\}  \notag
\end{align}%
Here we must replace the Bogoliubov coefficients eqs. (\ref{bog-sw}) and (%
\ref{bog-lw}). We are interested in short wavelength velocity modes, i.e.,
those inside the particle horizon for which $q\geq 1$. However, care must be
taken when $\bar{\kappa}$ approaches $\bar{\varsigma}$ as in this case the
Bogoliubov for long wavelengths must be used. After long but straightforward
calculations we obtain that the full expressions for the contribution of
long wavelengths to the velocity spectrum is

\begin{align}
\Phi _{(1)\left( l\right) }^{ij}\left( \varsigma ,\tau \right) & \simeq 
\frac{\pi }{16}\left( \frac{9}{16}\right) ^{4}\frac{1}{\tau _{0}^{2}}\left( 
\frac{H}{m}\right) ^{2}\frac{H^{5}}{32\pi ^{3}a^{4}\left( \tau \right) }%
\frac{1}{\left\langle \rho +p\right\rangle ^{2}}\frac{\varsigma
^{i}\varsigma ^{j}}{\varsigma ^{13}}  \label{corr-22-e} \\
& \times \left[ \frac{3}{2}a^{2}\left( \tau \right) +\frac{1}{16a^{5}\left(
\tau \right) }+\frac{1}{8a^{2}\left( \tau \right) }+\frac{3}{2a\left( \tau
\right) }-\frac{3}{2a\left( \tau \right) }\left( \frac{am}{H}\right)
^{m^{2}/H^{2}}\right]  \notag
\end{align}%
\begin{align}
\Phi _{(2)\left( 0\right) }^{ij}\left( \varsigma ,\tau \right) & \simeq 
\frac{1}{16}\left( \frac{9}{16}\right) ^{4}\frac{1}{\tau _{0}^{2}}\frac{H^{5}%
}{32\pi ^{2}a^{4}\left( \tau \right) }\frac{1}{\left\langle \rho
+p\right\rangle ^{2}}\frac{\delta ^{ij}}{6\varsigma ^{11}}  \label{corr-23-e}
\\
& \times \left[ 1+3\left( \frac{H}{m}\right) ^{2}\left( 1-\left[ \frac{%
a\left( \tau \right) m}{H}\right] ^{m^{2}/H^{2}}\right) \right]  \notag
\end{align}%
while for short wavelengths we have%
\begin{align}
\Phi _{(1)\left( s\right) }^{ij}&\left( \varsigma ,\tau \right)  \simeq 
\notag\\
&\simeq \frac{H^{5}}{8\pi ^{2}a^{3}\left( \tau \right) }\frac{1}{\tau _{0}^{4}}
\frac{%
1}{\left\langle \rho +p\right\rangle ^{2}}\left( \frac{9}{16}\right) ^{8}%
\frac{1}{\left( 16\right) ^{2}}\frac{1}{11}\frac{19!}{9!}\left[ \frac{1}{%
10!10}-\sum_{n=1}^{9}\frac{\left( 9-n\right) !}{\left( 20-n\right) !}\right] 
\frac{\varsigma ^{i}\varsigma ^{j}}{\varsigma ^{11}}  \notag \\
& +\frac{3H^{5}}{32\pi ^{2}a^{4}\left( \tau \right) }\left( \frac{H}{m}%
\right) ^{4}\frac{1}{\tau _{0}^{2}}\frac{1}{\left\langle \rho
+p\right\rangle ^{2}}\frac{1}{4}\left( \frac{9}{16}\right) ^{4}\frac{1}{16}%
\frac{\varsigma ^{i}\varsigma ^{j}}{\varsigma ^{11}}  \notag \\
& +\frac{H^{5}}{64\pi ^{2}a^{3}\left( \tau \right) }\frac{220}{3}\left[ 
\frac{1}{4}-\sum_{n=1}^{9}\frac{\left( 9-n\right) !}{\left( 12-n\right) !}%
\right] \left( \frac{9}{16}\right) ^{4}\frac{1}{16}\left( \frac{H}{m}\right)
^{2}\frac{1}{\tau _{0}^{2}}\frac{1}{\left\langle \rho +p\right\rangle ^{2}}%
\frac{\varsigma ^{i}\varsigma ^{j}}{\varsigma ^{11}}  \notag \\
& \simeq \frac{3H^{5}}{128\pi ^{2}a^{4}\left( \tau \right) }\left( \frac{9}{%
16}\right) ^{4}\frac{1}{16}\left( \frac{H}{m}\right) ^{4}\frac{1}{\tau
_{0}^{2}}\frac{1}{\left\langle \rho +p\right\rangle ^{2}}\frac{\varsigma
^{i}\varsigma ^{j}}{\varsigma ^{11}}  \label{corr-28-d}
\end{align}%
\begin{align}
\Phi _{(2)\left( s\right) }^{ij}&\left( \varsigma ,\tau \right)  \simeq
\notag\\ 
&\simeq \frac{H^{5}}{8\pi ^{2}a^{3}\left( \tau \right) }\frac{1}{\tau _{0}^{4}}
\frac{%
1}{\left\langle \rho +p\right\rangle ^{2}}\left( \frac{9}{16}\right) ^{8}%
\frac{1}{\left( 16\right) ^{2}}\frac{1}{11}\frac{17!}{7!}\left[ \frac{1}{10}%
\frac{1}{10!}-\sum_{p=1}^{n-1}\frac{\left( 7-p\right) !}{\left( 18-p\right) !%
}\right] \frac{\delta ^{ij}}{3\varsigma ^{9}}  \notag \\
& +\left( \frac{9}{16}\right) ^{4}\frac{1}{64}\frac{H^{5}}{64\pi
^{2}a^{4}\left( \tau \right) }\left( \frac{H}{m}\right) ^{4}\frac{1}{\tau
_{0}^{4}}\frac{1}{\left\langle \rho +p\right\rangle ^{2}}\frac{\delta ^{ij}}{%
\varsigma ^{9}}  \notag \\
& +\frac{H^{5}}{128\pi ^{2}a^{3}\left( \tau \right) }\left( \frac{H}{m}%
\right) ^{2}\frac{1}{\tau _{0}^{2}}\frac{1}{\left\langle \rho
+p\right\rangle ^{2}}\left( \frac{9}{16}\right) ^{4}\frac{72}{16}\frac{4}{3}%
\left[ \frac{1}{4}-\sum_{n=1}^{7}\frac{\left( 7-n\right) !}{\left(
10-n\right) !}\right] \frac{\delta ^{ij}}{3\varsigma ^{9}}  \notag \\
& \simeq \left( \frac{9}{16}\right) ^{4}\frac{1}{16}\frac{H^{5}}{128\pi
^{2}a^{4}\left( \tau \right) }\left( \frac{H}{m}\right) ^{4}\frac{1}{\tau
_{0}^{2}}\frac{1}{\left\langle \rho +p\right\rangle ^{2}}\frac{\delta ^{ij}}{%
\varsigma ^{9}}  \label{corr-29-d}
\end{align}

As $\varsigma >1$ the main contribution comes from term whose inverse power
of $\varsigma $ is the smallest, so we shall keep only them. Observe that
they come from the contribution of short wavelengths, and therefore depends
strongly on the details of the transition inflation-reheating. The two
contributions to the velocity spectrum are then%
\begin{equation}
\Phi _{(1)}^{ij}\left( \varsigma ,\tau \right) \simeq \left( \frac{9}{16}%
\right) ^{4}\frac{3}{16\times 128}\frac{H^{5}}{\pi ^{2}a^{4}\left( \tau
\right) }\left( \frac{H}{m}\right) ^{4}\frac{1}{\tau _{0}^{2}}\frac{1}{%
\left\langle \rho +p\right\rangle ^{2}}\frac{\varsigma ^{i}\varsigma ^{j}}{%
\varsigma ^{11}}  \label{corr-vel-1}
\end{equation}%
\begin{equation}
\Phi _{(2)}^{ij}\left( \varsigma ,\tau \right) \simeq \left( \frac{9}{16}%
\right) ^{4}\frac{1}{16\times 128}\frac{H^{5}}{\pi ^{2}a^{4}\left( \tau
\right) }\left( \frac{H}{m}\right) ^{4}\frac{1}{\tau _{0}^{2}}\frac{1}{%
\left\langle \rho +p\right\rangle ^{2}}\frac{\delta ^{ij}}{\varsigma ^{9}}
\label{corr-vel-2}
\end{equation}%
Replacing eq. (\ref{rho-p}) we obtain eq. (\ref{corr-9-b}).

\section{Calculation of the kinetic helicity}

We start from eq. (\ref{kin-hel-5}) and replace decomposition (\ref{jb}) of
the fields, thus obtaining%
\begin{align}
\Xi _{c}^{\Phi }\left( \tau ,\bar{x},\bar{x}^{\prime }\right) & =\frac{%
4H^{18}}{\left( 2\pi \right) ^{18}16\left\langle \rho +p\right\rangle
^{4}a^{30}\left( \tau \right) }\int d\bar{\varpi}d\bar{\varsigma}d\bar{%
\varkappa}d\bar{\sigma}d\bar{\varpi}^{\prime }d\bar{\varsigma}^{\prime }d%
\bar{\sigma}^{\prime }d\bar{\kappa}^{\prime }\notag\\
&\times \epsilon ^{ijk}\kappa
^{i}\varsigma ^{j}\varpi ^{k}\epsilon ^{ijk}\kappa ^{\prime i}\varsigma
^{\prime j}\varpi ^{\prime k}  \notag \\
& \times \left\{ e^{i\left( \bar{\kappa}+\bar{\varpi}+\bar{\varsigma}+\bar{%
\sigma}\right) \cdot \bar{x}}e^{-i\left( \bar{\kappa}^{\prime }+\bar{\varpi}%
^{\prime }+\bar{\varsigma}^{\prime }+\bar{\sigma}^{\prime }\right) \cdot 
\bar{x}^{\prime }}\dot{\phi}_{\sigma ^{\prime }}^{I\ast }\dot{\phi}%
_{\varsigma ^{\prime }}^{I\ast }\phi _{\kappa ^{\prime }}^{I\ast }\phi
_{\varpi ^{\prime }}^{I\ast }\dot{\phi}_{\sigma }^{I}\dot{\phi}_{\varsigma
}^{I}\phi _{\kappa }^{I}\phi _{\varpi }^{I}\right.   \label{ap4-2} \\
& \times \left[ a_{\sigma }b_{\varsigma }b_{k}a_{\varpi }-b_{\sigma
}a_{\varsigma }b_{\kappa }a_{\varpi }\right] \left[ a_{\sigma ^{\prime
}}^{\dagger }b_{\varsigma ^{\prime }}^{\dagger }b_{\kappa ^{\prime
}}^{\dagger }a_{\varpi ^{\prime }}^{\dagger }-b_{\sigma ^{\prime }}^{\dagger
}a_{\varsigma ^{\prime }}^{\dagger }b_{\kappa ^{\prime }}^{\dagger
}a_{\varpi ^{\prime }}^{\dagger }\right]   \notag \\
& +e^{i\left( \bar{\kappa}^{\prime }+\bar{\varpi}^{\prime }+\bar{\varsigma}%
^{\prime }+\bar{\sigma}^{\prime }\right) \cdot \bar{x}\prime }e^{-i\left( 
\bar{\kappa}+\bar{\varpi}+\bar{\varsigma}+\bar{\sigma}\right) \cdot \bar{x}}%
\dot{\phi}_{\sigma }^{I\ast }\dot{\phi}_{\varsigma }^{I\ast }\phi _{\kappa
}^{I\ast }\phi _{\varpi }^{I\ast }\dot{\phi}_{\sigma ^{\prime }}^{I}\dot{\phi%
}_{\varsigma ^{\prime }}^{I}\phi _{\kappa ^{\prime }}^{I}\phi _{\varpi
^{\prime }}^{I}  \notag \\
& \times \left. \left[ a_{\sigma ^{\prime }}b_{\varsigma ^{\prime
}}b_{\kappa ^{\prime }}a_{\varpi ^{\prime }}-b_{\sigma ^{\prime
}}a_{\varsigma ^{\prime }}b_{\kappa ^{\prime }}a_{\varpi ^{\prime }}\right] %
\left[ a_{\sigma }^{\dagger }b_{\varsigma }^{\dagger }b_{\kappa }^{\dagger
}a_{\varpi }^{\dagger }-b_{\sigma }^{\dagger }a_{\varsigma }^{\dagger
}b_{\kappa }^{\dagger }a_{\varpi }^{\dagger }\right] \right\}   \notag
\end{align}%
Noting that to avoid an odd integrand, we must contract$\ \sigma ^{\prime }$
with $\sigma $, we are left with%
\begin{align}
\Xi _{c}^{\Phi }\left( \tau ,\bar{x},\bar{x}^{\prime }\right) & =\frac{%
4H^{18}}{\left( 2\pi \right) ^{33/2}16\left\langle \rho +p\right\rangle
^{4}a^{27}\left( \tau \right) }\int d\bar{\sigma}\dot{\phi}_{\sigma
}^{I}\left( \tau \right) \dot{\phi}_{\sigma }^{I\ast }\left( \tau ^{\prime
}\right)   \notag \\
& \times \int d\bar{\varpi}d\bar{\varsigma}d\bar{\kappa}d\bar{\varpi}%
^{\prime }d\bar{\varsigma}^{\prime }d\bar{\kappa}^{\prime }\epsilon
^{ijk}k^{i}\varsigma ^{j}\varpi ^{k}\epsilon ^{ijk}\kappa ^{\prime
i}\varsigma ^{\prime j}\varpi ^{\prime k}  \notag\\
&\times  e^{i\left( \bar{\kappa}+\bar{\varpi}+\bar{\varsigma}+\bar{\sigma}%
\right) \cdot \bar{x}}e^{-i\left( \bar{\kappa}^{\prime }+\bar{\varpi}%
^{\prime }+\bar{\varsigma}^{\prime }+\bar{\sigma}\right) \cdot \bar{x}%
^{\prime }}\notag \\
& \times\dot{\phi}_{\varsigma ^{\prime }}^{I\ast }\left( \tau ^{\prime
}\right) \phi _{\kappa ^{\prime }}^{I\ast }\left( \tau ^{\prime }\right)
\phi _{\varpi ^{\prime }}^{I\ast }\left( \tau ^{\prime }\right) \dot{\phi}%
_{\varsigma }^{I}\left( \tau \right) \phi _{\kappa }^{I}\left( \tau \right)
\phi _{\varpi }^{I}\left( \tau \right) a_{\varpi }b_{\kappa }  \label{ap4-3}
\\
& \times \left[ b_{\varsigma }b_{\varsigma ^{\prime }}^{\dagger
}+a_{\varsigma }a_{\varsigma ^{\prime }}^{\dagger }\right] b_{\kappa
^{\prime }}^{\dagger }a_{\varpi ^{\prime }}^{\dagger }\notag\\
&+\bar{x}\leftrightarrow \bar{x}^{\prime }  \notag
\end{align}%
where we used $\left[ a_{\kappa },a_{\varpi }^{\dagger }\right] =\left( 2\pi
\right) ^{3/2}a^{3}\left( \tau \right) \delta \left( \bar{\kappa}-\bar{\varpi%
}\right) $. Considering all possible combinations of the remaining moments,
we finally have
\begin{align}
\Xi _{c}^{\Phi }\left( \tau ,\bar{\xi}\right) & =\frac{4H^{18}}{16\left(
2\pi \right) ^{12}\left\langle \rho +p\right\rangle ^{4}a^{18}\left( \tau
\right) }\int d\bar{\sigma}d\bar{\varpi}d\bar{\varsigma}d\bar{\kappa}
\notag\\
& \times e^{i\left( \bar{\kappa}+\bar{\varpi}+\bar{\varsigma}
+\bar{\sigma}\right) \cdot \bar{\xi}} \left(\epsilon ^{ijk}\kappa ^{i}
\varsigma ^{j}\varpi ^{k}\right) ^{2} 
\left\vert \dot{\phi}_{q}^{I}\left( \tau \right) \right\vert
^{2}\left\vert \phi _{\varpi }^{I}\left( \tau \right) \right\vert ^{2} 
\label{ap4-5} \\
& \left[ 
\dot{\phi}_{\varsigma }^{I}\left( \tau \right) \dot{\phi}_{\varsigma
}^{I\ast }\left( \tau \right) \phi _{\kappa }^{I}\left( \tau \right) \phi
_{\kappa }^{I\ast }\left( \tau \right) -\phi _{\kappa }^{I}\left( \tau
\right) \dot{\phi}_{\kappa }^{I\ast }\left( \tau \right) \dot{\phi}%
_{\varsigma }^{I}\left( \tau \right) \phi _{\varsigma }^{I\ast }\left( \tau
\right) \right]   \notag \\
& +\bar{\xi}\leftrightarrow -\bar{\xi}  \notag
\end{align}%
with $\bar{\xi}=\bar{x}-\bar{x}^{\prime }$. Replacing $\phi _{k}^{I}\left(
\tau \right) =\alpha _{k}\phi _{k}^{R}\left( \tau \right) +\beta _{k}\phi
_{k}^{R\ast }\left( \tau \right) $ we obtain, as in the case of the velocity
correlation $R_{ij}$, several kernels: one with only the vacuum
contribution, another with mixed contributions from vacuum and from the
created particles, and a third one with the contribution of only the created
particles. The expressions are rather long, but they are straightforwardly
obtained . Of the one due to the created particles, the part with $%
\left\vert \beta _{k}\right\vert ^{2}$ gives the main contribution, because
as was the case for $R_{ij}$, terms with $\alpha _{\kappa }\alpha _{\varpi
}^{\ast }\ldots $, etc. oscillate, and will give negligible contributions
when integrated. Therefore we shall consider 
\begin{align}
\Xi_c^{\Phi} \left( \tau ,\bar{\xi}\right) & \simeq \frac{H^{18}}{8\left( 2\pi
\right) ^{12}\left\langle \rho +p\right\rangle ^{4}a^{18}\left( \tau \right) 
}\int d\bar{\sigma}d\bar{\varpi}d\bar{\varsigma}d\bar{\kappa}\left\vert
\beta _{\sigma }\right\vert ^{2}\left\vert \beta _{\varpi }\right\vert
^{2}\left\vert \beta _{\varsigma }\right\vert ^{2}\left\vert \beta _{\kappa
}\right\vert ^{2}  \notag \\
& \times e^{i\left( \bar{\kappa}+\bar{\varpi}+\bar{\varsigma}+\bar{%
\sigma}\right) \cdot \bar{\xi}}\left( \epsilon ^{ijk}\kappa ^{i}
\varsigma ^{j}\varpi^{k}\right) ^{2} 
\left\vert \dot{\phi}_{\sigma }^{R}\left( \tau \right) \right\vert
^{2}\left\vert \phi _{\varpi }^{R}\left( \tau \right) \right\vert
^{2} \label{ap4-6}\\
&\times \left\{ \dot{\phi}_{\varsigma }^{R}\left( \tau \right) \dot{\phi}%
_{\varsigma }^{R\ast }\left( \tau \right) \phi _{\kappa }^{R}\left( \tau
\right) \phi _{\kappa }^{R\ast }\left( \tau \right) -\phi _{\kappa
}^{R}\left( \tau \right) \dot{\phi}_{\kappa }^{R\ast }\left( \tau \right) 
\dot{\phi}_{\varsigma }\left( \tau \right) \phi _{\varsigma }^{R\ast }\left(
\tau \right) \right\}   \notag \\
& +\bar{\xi}\leftrightarrow -\bar{\xi}  \notag
\end{align}%
Replacing the WKB form for the modes and keeping only the slowly decaying
terms we can express eq. (\ref{ap4-6}) as%
\begin{align}
\Xi_c^{\Phi} \left( \tau ,\bar{x},\bar{x}^{\prime }\right) & \simeq \frac{H^{18}}{%
\left( 2\pi \right) ^{12}8\left\langle \rho +p\right\rangle ^{4}a^{18}\left(
\tau \right) }\int d\bar{\sigma}d\bar{\varpi}d\bar{\varsigma}d\bar{\kappa}%
\left\vert \beta _{\sigma }\right\vert ^{2}\left\vert \beta _{\varpi
}\right\vert ^{2}\left\vert \beta _{\varsigma }\right\vert ^{2}\left\vert
\beta _{\kappa }\right\vert ^{2}  \notag \\
& \times e^{i\left( \bar{\kappa}+\bar{\varpi}+\bar{
\varsigma}+\bar{\sigma}\right) \cdot \bar{\xi}}
\left( \epsilon ^{ijk}\kappa ^{i}\varsigma
^{j}\varpi ^{k}\right) ^{2}
\frac{\Omega _{\sigma }
\left( \tau \right) }{\Omega _{\varpi
}\left( \tau \right) }\left[ \frac{\Omega _{\varsigma }\left( \tau \right) }{%
\Omega _{\kappa }\left( \tau \right) }-1\right] \label{ap4-7}\\
&+\bar{\xi}\left.\leftrightarrow \right. -\bar{\xi}  \notag
\end{align}%
Working in spherical coordinates and performing the angular integrals for
each mode we are left with%
\begin{eqnarray}
\Xi_c^{\Phi} \left( \tau ,\bar{\xi}\right)_c^{} &\simeq& -\frac{4H^{18}}{\left( 2\pi \right)
^{8}\left\langle \rho +p\right\rangle ^{4}a^{18}\left( \tau \right) }\frac{1%
}{\xi }I_{1}\left( \xi \right) \frac{\partial ^{2}}{\partial \xi ^{2}}\left[ 
\frac{1}{\xi }I_{2}\left( \xi \right) \right] \label{ap4-8}\\
&\times&\left\{ \frac{\partial ^{2}}{%
\partial \xi ^{2}}\left[ \frac{1}{\xi }I_{1}\left( \xi \right) \right] \frac{%
\partial ^{2}}{\partial \xi ^{2}}\left[ \frac{1}{\xi }I_{2}\left( \xi
\right) \right] -\left( \frac{\partial ^{2}}{\partial \xi ^{2}}\left[ \frac{1%
}{\xi }I_{3}\left( \xi \right) \right] \right) ^{2}\right\}   \nonumber
\end{eqnarray}
where%
\begin{align}
I_{1}\left( \xi \right) & =\int d\kappa \kappa \sin \left( \kappa \xi
\right) \left\vert \beta _{\kappa }\right\vert ^{2}\Omega _{\kappa }\left(
\tau \right)   \label{ap4-9a} \\
I_{2}\left( \xi \right) & =\int d\kappa \kappa \sin \left( \kappa \xi
\right) \left\vert \beta _{\kappa }\right\vert ^{2}\frac{1}{\Omega _{\kappa
}\left( \tau \right) }  \label{ap4-9b} \\
I_{3}\left( \xi \right) & =\int d\kappa \kappa \sin \left( \kappa \xi
\right) \left\vert \beta _{\kappa }\right\vert ^{2}  \label{ap4-9c}
\end{align}%
These integrals can be performed straighforwardly using the same
approximations as for $\Phi ^{ij}$, obtaining%
\begin{align}
\frac{1}{\xi }I_{1}\left( \xi \right) &\simeq \frac{1}{4}\frac{3H^{2}}{2m^{2}}
\left[ \cos \left( \xi \right) -1\right] +\left( \frac{9}{16}\right) ^{4}%
\frac{1}{16}\frac{1}{\tau _{0}^{2}}\notag\\
&\times \left\{ \frac{1}{7!}\left[ 6!-4!\xi
^{2}+2!\xi ^{4}-\xi ^{6}\right] \frac{\sin \left( \xi \right) }{\xi }\right.\notag \\
&-\left.\frac{1}{7!}\left[ -5!+3!\xi ^{2}-\xi ^{4}\right] \cos \left( \xi \right) +\frac{1!
}{7!}\xi ^{6}{\rm Ci}\left( \xi \right) \right\}   \label{ap4-9aa}
\end{align}
\begin{eqnarray}
\frac{1}{\xi }I_{2}\left( \xi \right) & \simeq& \frac{1}{4}\left( \frac{H}{m}%
\right) ^{2}\left[ -\frac{1}{\xi }\sin \left( \xi \right) +1+\frac{3H^{2}}{%
2m^{2}}\left[ \cos \left( \xi \right) -1\right] \right]  \nonumber\\
&-& \frac{H^{2}}{m^{2}}%
\frac{1}{\xi ^{2}}\left[ \cos \left( \xi \right) -1\right]
+ 1-\frac{1}{\xi } \sin \left( \xi \right) +\int_{m/H}^{1}dp\frac{\cos \left( p\xi \right) }{p} 
\nonumber\\
&+&\left( \frac{9}{16}\right) ^{4}\frac{1}{16}\frac{1}{\tau _{0}^{2}}\left\{ 
\frac{1}{9!}\left[ 8!-6!\xi ^{2}+4!\xi ^{4}-2!\xi ^{6}+\xi ^{8}\right] \frac{%
1}{\xi }\sin \left( \xi \right) \right.
\label{ap4-9bb} \\
&+&\left.\frac{1}{9!}\left[ 7!-5!\xi ^{2}+3!\xi
^{4}-\xi ^{6}\right] \cos \left( \xi \right) -\frac{1}{9!}\xi ^{8}{\rm Ci}%
\left( \xi \right) \right\}   \notag
\end{eqnarray}%
and%
\begin{align}
\frac{1}{\xi }I_{3}\left( \xi \right) & \simeq \frac{1}{4}\left\{ -\frac{H}{m%
}\frac{\sin \left( \xi \right) }{\xi }+\frac{H}{m}-\frac{1}{2}\frac{\sin
\left( \xi \right) }{\xi }+\frac{1}{2}\frac{H^{2}}{m^{2}}\frac{1}{\xi }\sin
\left( \frac{m}{H}\xi \right) \right. \notag\\
&- \left.\frac{1}{2}\frac{H}{m}\sin \left( \frac{m}{H}%
\xi \right) +\frac{1}{2}\xi \right\}   \label{ap4-9cc} \\
& +\left( \frac{9}{16}\right) ^{4}\frac{1}{16}\frac{1}{8!}\frac{1}{\tau
_{0}^{2}}\left\{ \left[ 7!-5!\xi ^{2}+3!\xi ^{4}-\xi ^{6}\right] \frac{\sin
\left( \xi \right) }{\xi }\right. \notag\\
&+\left.\left[ 6!-4!\xi ^{2}+2!\xi ^{4}-\xi ^{6}\right]
\cos \left( \xi \right) +\xi ^{7}\left[ \frac{\pi }{2}-{\rm Si}\left( \xi
\right) \right] \right\}   \notag
\end{align}

\end{document}